\documentclass[12pt%
]{article}

\usepackage[margin=1in]{geometry}

\usepackage{setspace}
\usepackage{latexsym}
\usepackage{amssymb,amsmath, bm}
\usepackage{graphicx}
\usepackage{subcaption}  
\usepackage{caption}
\usepackage{marvosym}

\usepackage{tikz}
\usetikzlibrary{shapes,arrows}
\usetikzlibrary{positioning}

\usepackage{natbib}
\usepackage{color}
\usepackage[hidelinks, pdftex, bookmarksopen=true, bookmarksnumbered=true,
pdfstartview=FitH, breaklinks=true, urlbordercolor={0 1 0}, citebordercolor={0 0 1}]{hyperref}
\definecolor{darkgreen}{rgb}{0,0.545,0}
\definecolor{darkyellow}{rgb}{0.933,0.604,0}

\usepackage{longtable}
\usepackage{booktabs}

\usepackage{dcolumn}
\newcolumntype{.}{D{.}{.}{-1}}
\newcolumntype{d}[1]{D{.}{.}{#1}}

\usepackage{rotating}

\usepackage{bbm}

\newcommand{\blind}{0}

\usepackage{theorem}
\theoremstyle{break}
\theoremheaderfont{\scshape}
\theorembodyfont{\rm}

\newcommand\addprime{^{\prime}}
\newcommand\addtop{^{\top}}

\newcommand\iid{\stackrel{\sf i.i.d.}{\sim}}

\renewcommand\r{\right}
\renewcommand\l{\left}

\newcommand\bbeta{{\boldsymbol \beta}}

\usepackage{arydshln}

\usepackage[compact]{titlesec}

\usepackage{xr}
\allowdisplaybreaks[4]

\begin{document}

\newcommand\spacingset[1]{\renewcommand{\baselinestretch}%
{#1}\small\normalsize}

\spacingset{1.1}


\newcommand{\titbase}{\bf A Dynamic Dirichlet Process Mixture Model for the Partisan Realignment of Civil Rights Issues in the U.S.~House of Representatives}

\newcommand{\tit}{\titbase}

\if0\blind
{
\title{\tit\thanks{
\protect We thank Eric Schickler for generously providing the data used in our analysis. We also thank Dean Knox, Christopher Lucas, Jacob Montgomery, Iain Osgood, Mark Pickup, Kevin Quinn, and audiences at the 2020 Winter Meeting of Japanese Society for Quantitative Political Science, the 37th Annual Summer Meeting of the Society for Political Methodology, and Virtual Speaker Series in the Department of Political Science at Washington University in St. Louis for insightful comments and suggestions.

}
}

\author{
 Nuannuan Xiang\thanks{
    Assistant Professor in Health Policy and Management, Mailman School of Public Health, Columbia University.
    Email:~\href{mailto:nx2174@cumc.columbia.edu}{\tt nx2174@cumc.columbia.edu}.
    }
    \and
 Yuki Shiraito\thanks{
    Assistant Professor, Department of Political Science, University of Michigan.
    URL:~\href{https://shiraito.github.io}{\tt shiraito.github.io}.
    }
    \thanks{
    Correspondence: Center for Political Studies, 4259 Institute for Social Research, 426 Thompson Street, Ann Arbor, MI 48104-2321.  Phone: 734-615-5165, Email: \href{mailto:shiraito@umich.edu}{\tt shiraito@umich.edu}.
    }
    }
\date{
This draft: February 12, 2025 \\
First draft: July 7, 2020
}

}\fi

\if1\blind
{
\title{\bf \tit}
}\fi

\maketitle

\pdfbookmark[1]{Title Page}{Title Page}

\thispagestyle{empty}
\setcounter{page}{0}

 \begin{abstract}
  Evolutionary societal changes often prompt a debate.
The positions of the two major political parties in the United States on civil rights issues underwent a reversal in the 20th century.
The conventional view holds that this shift was a structural break in the 1960s, driven by party elites, while recent studies argue that the change was a more gradual process that began as early as the 1930s, driven by local rank-and-file party members.
Motivated by this controversy, this paper develops a nonparametric Bayesian model that incorporates a hidden Markov model into the Dirichlet process mixture model.
A distinctive feature of the proposed approach is that it models a process in which multiple latent clusters emerge and diminish as a continuing process so that it uncovers any of steady, sudden, and repeated shifts in analysing longitudinal data.
Our model estimates each party's positions on civil rights in each state based on the legislative activities of their Congressional members, identifying cross- and within-party coalitions over time.
We find evidence of gradual racial realignment in the 20th century, with two periods of fast changes during the 1948 election and the Civil Rights Movement.



 \end{abstract}

\clearpage


\onehalfspacing

\section{Introduction}
Gradual structural changes---changes that remain obscure while they are occurring but become clear after a few decades---often generate debates among social scientists.
Since the process of changes is not directly observable, researchers need to infer when the latent shift began and how it evolved.
Political scientists studying party politics in the United States are well aware that the Democratic and Republican parties switched their positions on racial issues during the 20th century.
At the end of the 20th century, the Democratic Party is associated with racial liberalism, while the Republican Party is linked to racial conservatism.
However, the positions of these parties were the opposite in the 1930s.
The conventional wisdom often frames this shift as a structural break, in which party elites in Washington, D.C. orchestrated a sudden reshuffling of the party positions in the 1960s \citep{Carmines_Stimson_1989}.
Recent studies challenge this view, arguing that the behaviour of rank-and-file legislators began to change gradually in the 1940s, or even as early as the 1930s \citep{Schickle16, Schickler_et_al_10, chen_2009, ChenMickeyHouweling}.

We revisit this debate by developing a dynamic model of gradual structural changes that extends the Dirichlet process (DP) mixture model \citep{ferguson1973bayesian,antoniak1974mixtures} to reanalyse historical data.
The DP mixture model is a nonparametric Bayesian model for clustering units into latent groups.
It places a DP prior on the mixing distribution of a mixture model, allowing the number of clusters to be estimated from data, instead of requiring it to be specified \emph{ex ante}.
We extend the DP mixture model to a dynamic setting by modifying the standard DP prior to a Markov process of the DP for changing assignment of units to clusters over time.
Like the static DP, this dynamic DP is closely related to the Chinese restaurant process \citep{blackwell1973ferguson}.
We call it the intergenerational Chinese restaurant process because the cluster assignments in each time period are conditioned on the cluster assignments from the previous period.

A distinctive feature of our model is that it captures a process in which multiple clusters emerge and diminish gradually, rather than a one-time structural break.
At the same time, our model retains the key feature of the DP mixture model that the number of clusters does not need to be specified \emph{ex ante}.
Moreover, as with the DP mixture model, a broad class of models can serve as the mixture component in the dynamic DP mixture model, making it widely applicable beyond our specific focus on party position switching on racial issues.

We use the dynamic DP mixture model to analyse a dataset used by Eric Schickler in his book \textit{Racial Realignment: The Transformation of American Liberalism, 1932-1955}.\footnote{This dataset is also featured in \citet{Schickler_et_al_10}. Schickler also analyses other types of data, such as survey data and state party platforms, in his book.} This dataset includes four types of legislative activities on civil rights issues---roll-call votes, petition signatures for discharging bills, floor speeches, and bill sponsorships respectively---in the House of Representatives spanning the 73rd to the 92nd Congress (1933--1973).

In our analysis, we treat each party in each state as a unit of analysis, identifying latent clusters among these state-party units based on the legislative activities of their Congressional members. The dynamic DP mixture model enables us to infer gradual changes in their clustering pattern over time, thereby answering the question of when and how the two parties in each state changed their positions on racial issues.

Compared to the original analysis by \citet{Schickle16}, our approach offers the advantage of analysing all types of legislative activities simultaneously, which allows us to leverage a broader set of information and reveal a clearer pattern in how the two parties in each state shifted and realigned their positions. Consistent with Schickler's argument, we observe a distinct trend in the 1940s, in which the two parties in Northern states quickly diverged on civil rights issues. Additionally, we find that since the 1950s, the two parties in each Southern state tended to adopt similar stances on civil rights, a pattern that persisted until the pivotal shifts of the national leaders' positions around 1965 suddenly disrupted this regional alignment between the two parties in the South. We also find that the solidarity of Southern Democrats declined quickly during the Civil Rights Movement.

Our extension of the DP mixture model is characterised by the dependence of cluster assignments on the previous period.
The methodological challenge of our application is to capture gradual changes in latent heterogeneity over time.
We use a mixture model to account for latent heterogeneity, which is common in political science research \citep{imai2012statistical,spirling2010identifying,10.1111/j.1467-9876.2011.00768.x}.
However, the change-point modelling approach, which is also common to model temporal dynamics in social sciences \citep{park10,Pang12:_endogenous_jurisprudential_regimes,kim:liao:imai:19} is not suitable due to its assumption that all units switch their status simultaneously and latent clusters are not shared across time.
In this paper, we develop a new dynamic DP prior characterised by ``stickiness'' of cluster assignments over time \citep{caron2012generalizedpolyaurntimevarying,Fox_et_al_2011}.

The remainder of the paper proceeds as follows. The next section introduces our motivating empirical example, racial realignment of the Democratic and Republican Parties. Section \ref{sec:model} first describes the proposed model using the stick-breaking definition of the Dirichlet process. Then the section illustrates the intergenerational Chinese restaurant metaphor for the proposed model, which provides the intuition on how latent clusters evolve over time.
We also discuss the differences and similarities between the proposed model and existing models in Section \ref{sec:model}.
Section \ref{sec:analysis} presents the empirical analysis of the motivating example, followed by concluding remarks.

\section{Motivating Example and Data}

\subsection{Racial Realignment, 1932-1965}
The two major parties in the United States swapped their positions on racial issues with each other during the 20th century.
As ``Lincoln's party'', the Republican Party used to be more closely associated with African Americans than the Democrats at the beginning of the century.
By the end of the century, however, the Democratic Party became associated with racial liberalism, advocating for government efforts to address racial inequality, while the Republicans became linked to racial conservatism, more likely to oppose governmental interventions in racial inequality.

Traditionally, scholars view the reversal in the two parties' positions on racial issues as a sudden structural break that occurred in the 1960s.
According to this view, although local Northern Democrats had an incentive to support civil rights to gain African American voters, who were becoming increasingly important in the North, federalism constrained local politicians from committing to programmatic liberalism \citep{weir2005}. Instead, party elites in Washington, D.C. led the change. The ``critical juncture'' arrived during the 1964 presidential election, when Democratic candidate Lyndon B. Johnson and Republican candidate Barry Goldwater took sharply opposing positions on civil rights \citep{Carmines_Stimson_1989,Califano1992,Edsall_Edsall_1992}.
Subsequently, local party activists aligned their racial positions with those of the national leaders.

By contrast, \citet{Schickle16} argues that Northern Democrats at the local level took the lead in driving this change.\footnote{Following \citet{Schickle16}, we define ``Southern'' states or the ``South'' as the 11 Confederate states plus Kentucky and Oklahoma. ``Northern'' states or the ``North'' refers to all other states.\label{footnote:north_south}} Local Northern Democrats were gradually transformed by the Democratic Party's New Deal coalition with the Congress of Industrial Organisations (CIO), African Americans, and other urban liberals in the North at the local level. Initially, the New Deal coalition, built on shared interests in economic liberalism, had little connection to civil rights advocacy. However, as the Democrats' nonpartisan allies increased their civil rights advocacy, Northern rank-and-file party members followed suit, advancing civil rights advocacy accordingly. While party elites in Washington, D.C. had a strong interest in maintaining the traditional North--South coalition within the Democratic Party, pressures from Northern rank-and-file members eventually forced national leaders to break that solidarity. Meanwhile, the Republican Party, which had previously been more supportive of civil rights, became increasingly divided over racial issues as the demand for civil rights waned in some of its local constituencies.

\citet{Schickle16} provides extensive data to support his argument, including historical survey data tracing the configuration of economic and racial liberalism at the individual level, records of state party platforms to identify when and where civil rights issues became (or ceased to be) significant concerns for the two parties at the state level, and data on House members' legislative activities to trace how rank-and-file Congress members---who were more responsible to their state-level constituencies than to party elites in Washington, D.C.---adjusted their racial positions to align with their constituencies' preferences.

In this paper, we reanalyse the legislative data mentioned above to understand how the state-level Democratic and Republican Parties, as represented by their Congress members, broke and remade coalitions on civil rights issues across the North-South divide as well as across party lines, which is a crucial dynamics discussed in \citet{Schickle16} but never directly analysed.

\subsection{Data}
The legislative data includes four types of activities in the House of Representatives: roll-call votes, signing petitions to discharge bills from committee, floor speeches, and bill sponsorships. Roll-call votes are the most commonly used data to measure Congress members' preferences \citep{poole1985spatial,clinton2004statistical}.
However, Schickler points out that roll-call data can be misleading because only a very small fraction of civil rights bills reached the floor---most were blocked by the House Rules Committee, which was dominated by senior southern Democrats \citep[p. 180-181]{Schickle16}. To address this limitation, Schickler uses data on petitions, speeches, and bill sponsorships to complement the roll-call measure.

Summary statistics for these four types of legislative data, covering the 73rd Congresses (1933--35) to the 92nd Congress (1971--73), are provided in Supplementary Information (SI) \ref{tab:datasummary}.

\textbf{Roll-call Votes.} The roll-call vote data consists of a series of dummy variables indicating whether a Congress member supported a civil rights bill: 1 for voting ``yes'' and 0 for voting ``no'' or being absent. There may be multiple roll calls on civil rights issues in a given Congress or none in another.

\textbf{Discharge Petitions.} If a legislative committee or the Rules Committee blocks a bill from reaching the floor for a long time, House members can sign a petition to advance the bill.\footnote{For a bill in a legislative committee, the waiting period is 20 days; for a bill in the Rules Committee, the waiting period is 7 days \citep[p. 676]{Schickler_et_al_10}. Schickler explains that signing a discharge petition is costly for members of Congress, because it is a sign of violating ``congressional norms'' and intruding on ``committee authority.'' Only members who ``cared enough'' about a civil rights issue would sign a petition \citep[p. 183]{Schickle16}.} The discharge petition data consists of a series of dummy variables indicating whether a Representative signed a particular petition to discharge a civil rights bill. Similar to roll-call votes, there may be no civil rights-related petitions or multiple such petitions in a given Congress.

\textbf{Floor Speeches.} Floor speeches are an important way for Congress members to signal commitments to their constituents. Competition for speech time on the House floor is intense, and using this limited time to deliver a speech in support of civil rights indicates that the issue was among the member's top priorities. The data on floor speeches consists of dummy variables indicating whether a Representative delivered at least one speech on the floor in support of civil rights during a given Congress.

\textbf{Bill Sponsorship.} Like floor speeches, bill sponsorship serves as a signal of strong support for a particular issue. The data on bill sponsorships consists of count variables indicating the number of civil rights bills a Representative sponsored in a given Congress.

\subsection{Limitations of the Original Analysis}
\citet{Schickle16} analyses the four types of data separately.\footnote{\citet{Schickler_et_al_10} also presents the results of analysing the same four types of data.}
While the analysis of roll-call votes shows that Northern Democrats were only slightly more likely than Northern Republicans to cast votes in support of civil rights before the 1960s, the analysis of discharge petitions reveals that, as early as the 1940s, Northern Democrats were substantially more likely to sign civil rights-related petitions than their Republican counterparts.
The newly discovered discharge petition data contribute significantly to the re-examination of racial realignment in the early twentieth century.
Previously, scholars relied on roll-call votes to trace changes in Congress members' positions on civil rights.
By using discharge petitions for measuring Congress members' preferences on civil rights issues, \citet{Schickle16} concludes that racial realignment began much earlier than the 1960s.
The analysis of speech data and bill sponsorship data further supports this argument.

Analysing the four types of data separately does not utilise the available information efficiently, and outright disregarding roll-call vote data risks introducing bias in favour of the book's argument. Instead, our model is flexible enough to analyse all four types of data simultaneously, allowing us to make full use of the observed legislative activities to infer the latent preferences of Congress members.

Moreover, \citet{Schickle16} discusses at length how rank-and-file party members allied across states and party lines in Congress but rarely directly evaluates this argument using data on legislative activities.
The original analysis focuses primarily on assessing the claim that Northern Democrats became more supportive of civil rights than the Republican Party well before the 1960s by examining differences in the four types of legislative activities between two parties in the North.
However, it does not address other arguments about coalition changes, such as the declining North--South coalition within the Democratic Party or the increasing cross-party coalition between Republicans and Southern Democrats.

Our model addresses the coalition question directly by tracing the evolving clustering patterns of state parties, as represented by their Congress members, over time. This approach allows us to examine how the Democratic Party's North--South coalition dissolved and how the cross-party coalition between Republicans and Southern Democrats emerged.


\section{Model and Inference} \label{model}
\label{sec:model}
\subsection{The Model}
The proposed dynamic DP mixture model is a nonparametric, dynamic Bayesian clustering model.

Like the DP mixture model, the dynamic DP mixture model identifies the latent cluster membership of units and is nonparametric in the sense that the number of latent clusters is not specified \textit{ex ante}.
However, while the DP mixture model clusters cross-sectional units, the dynamic DP mixture model is a model in which units move across clusters over time.
It is a dynamic model since  each unit's cluster assignment can change over time, and the temporal shift is modelled as a Markov process.
In each time period, a unit either remains in the same cluster as in the previous period or transitions to a different cluster. 
If a unit leaves its current cluster, the new cluster assignment is determined by a Dirichlet process conditional on the structure of cluster memberships in the previous period.
Therefore, the proposed model is a dynamic extension of the DP mixture model.

To formally describe the model, let $i \in \{1, \dots, N\}$ denote a unit and $t \in \{1, \dots, T\}$ denote a time period.
Also, let $Y_{it}$ be observed (possibly multivariate) measurement for unit $i$ in time $t$.
It is assumed that unit $i$ at time $t$ belongs to a latent cluster $g[it]$, where $g[it] \in \{1, 2, \dots \}$ denotes the latent cluster index of unit $i$ at time $t$, and that $Y_{it}$ is generated from a probabilistic model $f$ with parameter $\Theta_{g[it]}$:
\begin{align}
 Y_{it} \sim f \left( \Theta_{g[it]} \right).
 \label{eq:gen_dgp}
\end{align}
That is, $Y_{it}$ and $Y_{i\addprime t\addprime}$ share a common data generating process(DGP) if $g[it] = g[i\addprime t\addprime]$, but they may follow different DGPs otherwise.
An appropriate prior distributions are placed on parameters $\Theta$.
A simple example of $f$ is the linear regression model with cluster-specific parameters:
$$
Y_{it} \overset{\rm indep.} \sim \mathcal{N} \l(
X_{it}\addtop \bbeta_{g[it]}, \sigma^2_{g[it]} \r)
$$
In this example, $\Theta_{g} = \left( \bbeta_{g}, \sigma_{g} \right)$ and $\bbeta_{g}$ and $\sigma_{g}$ vary across latent clusters.

Now we turn to cluster assignments. First, cluster assignments in the initial period ($t = 1$) are generated by a regular DP.
Using the stick-breaking construction \citep{sethuraman1994constructive}, the generative process of $g[i1], i = 1, \dots, N,$ is hierarchically defined as:
\begin{align}
 g[i1] & \iid \textsf{Discrete} (\{q_{k}\}_{k=1}^{\infty})\label{eq:stkbrk_1} \\
 q_k & = \pi_k \prod_{l = 1}^{k-1}(1-\pi_l) \\
 \pi_k &\iid \textsf{Beta}(1,\gamma)\label{eq:stkbrk_3}
\end{align}
where $\gamma$ is the concentration parameter of the DP.

The key innovation of the proposed dynamic DP mixture model is the dynamic generative process of latent clusters for periods $t=2,\dots, T$.
Specifically, $g[it]$ follows a Markov process conditional on $(g[1, t- 1], \dots, g[N, t - 1])$.
On the one hand, unit $i$ in time $t$ stays in the same cluster as it was in time $t - 1$ with probability $p$, which we call the stickiness parameter \citep[cf.][]{Fox_et_al_2011}.
Formally,
\begin{align}
 \Pr \left( g[it] = g[i, t-1] \mid g[i, t-1] \right) &= p\label{eq:dynstkbrk_1} \\
 \intertext{where}
 p \sim \mathsf{Beta}(\alpha_p, \beta_p).
\end{align}
On the other hand, with probability $1-p$, unit $i$'s cluster assignment at time $t$ follows a DP, but this DP is conditional on the cluster assignments at $t - 1$.
That is,
\begin{align}
 g[it] & \iid \mathsf{Discrete}(\{q_k^t\}_{k = 1}^\infty) \\
 q_k^t & = \pi_k^t \prod_{l = 1}^{k-1}(1-\pi_l^t)\label{eq:dynstkbrk_q} \\
 \pi_k^t & \sim \mathsf{Beta}(1 + n_k^{t-1},\gamma + N - \sum_{l = 1}^k n^{t-1}_l)\label{eq:dynstkbrk_6}
\end{align}
where $n_{k}^{t - 1}$ is the number of units in latent cluster $k$ at $t - 1$.

\subsection{Intergenerational Chinese Restaurant Process Metaphor}
\label{subsec:igcrpmetaphor}
The Dirichlet process is known to have two constructive representations: the stick-breaking process and the Chinese restaurant process. Our dynamic DP mixture model, described above, is based on the stick-breaking process due to its mathematical simplicity. However, the Chinese restaurant process representation offers a more intuitive illustration of the DP. To further explain our model, we introduce a dynamic variation of the Chinese restaurant process, which we call the intergenerational Chinese restaurant process.
\begin{figure}[t!]
\centering
\begin{tikzpicture}
\path
( 0, 0) node [shape=circle,minimum size=2cm,draw] (table1) {$k = 1$}
( 4, 0) node [shape=circle,minimum size=2cm,draw] (table2){$k = 2$}

( 6.3, 0) node [shape=circle,draw,fill=black,minimum size =.01cm] {}
( 6.8, 0) node [shape=circle,draw,fill=black,minimum size =.01cm] {}
( 7.3, 0) node [shape=circle,draw,fill=black,minimum size =.01cm] {}

( -0.6,1.3) node [shape=rectangle,minimum size=0.5cm,draw] (obs1) {$i^{t=1} = 1$}
( 1, 3) node [shape=rectangle,minimum size=0.5cm,draw] (obs2_1){$i^{t=1} = 2$}

( 1, 1.4) node [color=black] {$\frac{1}{1 + \gamma}$}
( 3, 1.5) node [color=black] {$\frac{\gamma}{1 + \gamma}$};

\draw[->,very thick,color=black] (obs2_1) -- (table1);
\draw[->,very thick,color=black] (obs2_1) -- (table2);

\end{tikzpicture}

\vspace*{24pt}

\begin{tikzpicture}
\path
( 0, 0) node [shape=circle,minimum size=2cm,draw] (table1) {$k = 1$}
( 4, 0) node [shape=circle,minimum size=2cm,draw] (table2){$k = 2$}
( 8, 0) node [shape=circle,minimum size=2cm,draw] (table3){$k = 3$}

( 9.4, 0) node [shape=circle,draw,fill=black,minimum size =.01cm] {}
( 9.8, 0) node [shape=circle,draw,fill=black,minimum size =.01cm] {}
( 10.2, 0) node [shape=circle,draw,fill=black,minimum size =.01cm] {}
( -0.6,1.3) node [shape=rectangle,minimum size=0.5cm,draw] (obs1) {$i^{t=1} = 1$}

( 2.2, 0) node [shape=rectangle,minimum size=0.5cm,draw] (obs2){$i^{t=1} = 2$}
( 1, 3) node [shape=rectangle,minimum size=0.5cm,draw] (obs3_1) {$i^{t=1} = 3$}

( 1, 1.4) node [color=black] {$\frac{1}{2 + \gamma}$}
( 3.3, 1.5) node [color=black] {$\frac{1}{2 + \gamma}$}
( 6, 1.4) node [color=black] {$\frac{\gamma}{2 + \gamma}$};

\draw[->,very thick,color=black] (obs3_1) -- (table1);
\draw[->,very thick,color=black] (obs3_1) -- (table2);
\draw[->,very thick,color=black] (obs3_1) -- (table3);

\end{tikzpicture}

\vspace*{24pt}

\begin{tikzpicture}
\path
( 0, 0) node [shape=circle,minimum size=2cm,draw] (table1) {$k = 1$}
( 4, 0) node [shape=circle,minimum size=2cm,draw] (table2){$k = 2$}
( 8, 0) node [shape=circle,minimum size=2cm,draw] (table3){$k = 3$}
( 9.4, 0) node [shape=circle,draw,fill=black,minimum size =.01cm] {}
( 9.8, 0) node [shape=circle,draw,fill=black,minimum size =.01cm] {}
( 10.2, 0) node [shape=circle,draw,fill=black,minimum size =.01cm] {}
( -0.6,1.3) node [shape=rectangle,minimum size=0.5cm,draw] (obs1) {$i^{t=1} = 1$}
( 2.2, 0) node [shape=rectangle,minimum size=0.5cm,draw] (obs2){$i^{t=1} = 2$}
( 5.8, 0) node [shape=rectangle,minimum size=0.5cm,draw] (obs3) {$i^{t=1} = 3$}
( 1, 3) node [shape=rectangle,minimum size=0.5cm,draw] (obs4_1) {$i^{t=1} = 4$}
( 1, 1.4) node [color=black] {$\frac{1}{3 + \gamma}$}
( 3.3, 1.5) node [color=black] {$\frac{2}{3 + \gamma}$}
( 6, 1.4) node [color=black] {$\frac{\gamma}{3 + \gamma}$};
\draw[->,very thick,color=black] (obs4_1) -- (table1);
\draw[->,very thick,color=black] (obs4_1) -- (table2);
\draw[->,very thick,color=black] (obs4_1) -- (table3);
\end{tikzpicture}
\caption{\textbf{Chinese Restaurant Process.}
This figure illustrates the generative process of latent clusters at 
$t = 1$. For illustration purposes, we assume that in the realised states, unit 1 is assigned to cluster 1, unit 2 to cluster 2, and unit 3 to cluster 2. The fractions near the arrows represent the probabilities of units being assigned to the corresponding clusters. The probability of assigning a unit to an existing cluster is proportional to the number of units already assigned to that cluster. The probability of a unit creating a new cluster is proportional to the parameter $\gamma$. The top, middle, and bottom panels show the probabilities that units 2, 3, and 4 are assigned to each cluster, respectively.
}
\label{fig:crp}
\end{figure}
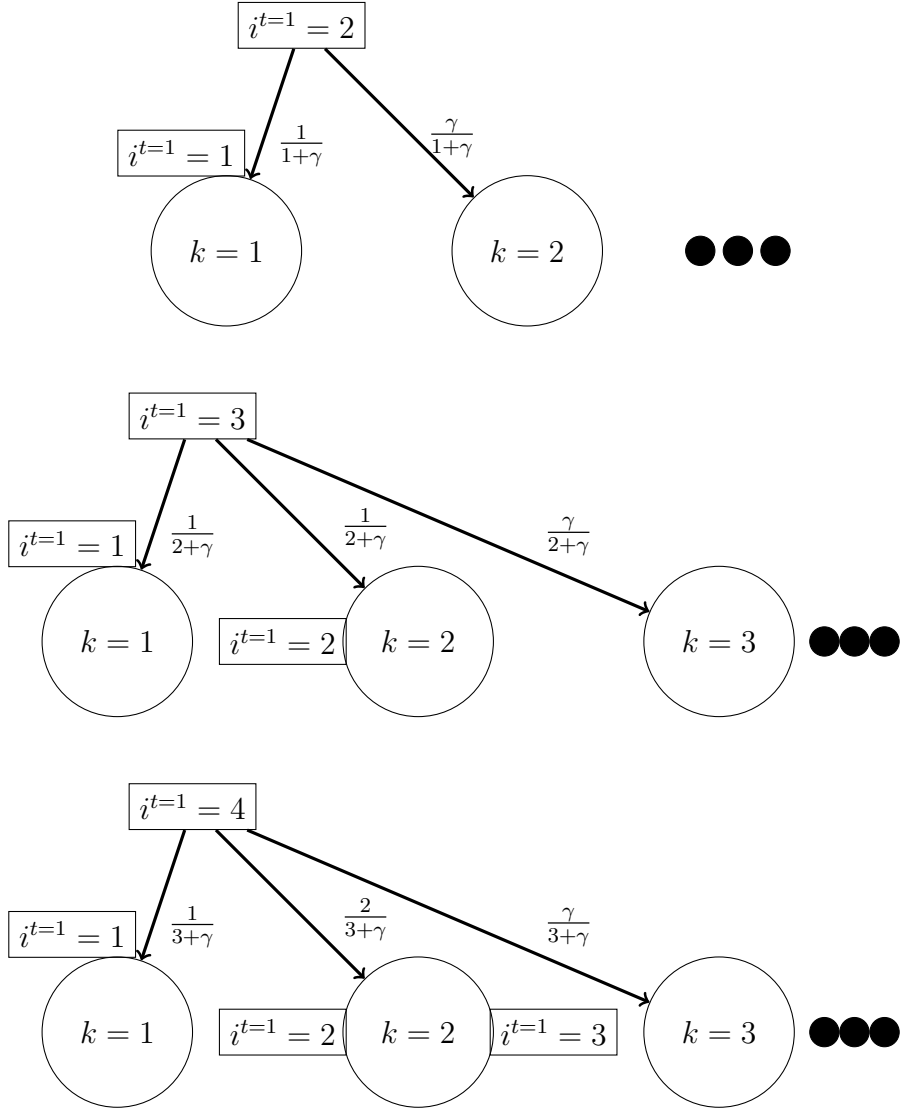

Our dynamic DP mixture model at $t = 1$ can be represented as a regular Chinese restaurant process, which is illustrated in Figure~\ref{fig:crp}. In this representation, units are assigned to latent clusters sequentially.
At the beginning, unit $1$ is (arbitrarily) assigned to cluster $1$.
For each subsequent unit, the probability that the unit is assigned to an existing cluster is proportional to the number of the units already assigned to that cluster, and the probability that the unit creates a new cluster is proportional to the concentration parameter $\gamma$.
For example, unit $2$ goes to cluster $1$ with probability $1 / (1 + \gamma)$ or creates a new cluster (cluster $2$) with probability $\gamma / (1 + \gamma)$ (the top panel of Figure~\ref{fig:crp}).
If unit $2$ is assigned to cluster $2$ in the realised state, the next unit (unit $3$) is assigned to cluster $1$ or $2$ with probability $1 / (2 + \gamma)$, or forms a new cluster (cluster $3$) with probability $\gamma / (2 + \gamma)$ (the middle panel of Figure \ref{fig:crp}).
Furthermore, given that unit $1$ is assigned to cluster $1$ and units $2$ and $3$ are assigned to cluster $2$, unit $4$ is assigned to cluster $1$ with probability $1 / (3 + \gamma)$, cluster $2$ with probability $2 / (3 + \gamma)$, and cluster $3$ with probability $\gamma / (3 + \gamma)$, respectively (the bottom panel of Figure~\ref{fig:crp}).
This process is equivalent to the model defined by equations~\eqref{eq:stkbrk_1} through \eqref{eq:stkbrk_3} for latent clusters at $t = 1$.

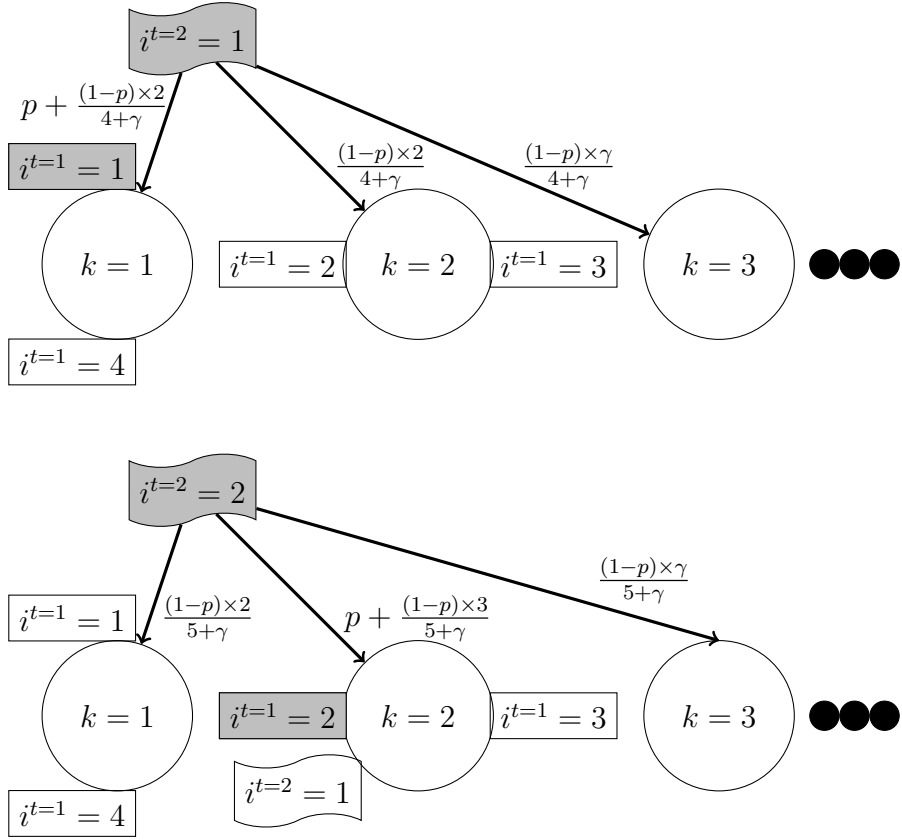
\begin{figure}[t]
\centering
\begin{tikzpicture}
\path
( 0, 0) node [shape=circle,minimum size=2cm,draw] (table1) {$k = 1$}
( 4, 0) node [shape=circle,minimum size=2cm,draw] (table2){$k = 2$}
( 8, 0) node [shape=circle,minimum size=2cm,draw] (table3){$k = 3$}
( 9.4, 0) node [shape=circle,draw,fill=black,minimum size =.01cm] {}
( 9.8, 0) node [shape=circle,draw,fill=black,minimum size =.01cm] {}
( 10.2, 0) node [shape=circle,draw,fill=black,minimum size =.01cm] {}

( -0.6,1.3) node [shape=rectangle,minimum size=0.5cm,draw] (obs1) {$i^{t=1} = 1$}
( 2.2, 0) node [shape=rectangle,minimum size=0.5cm,draw] (obs2_1){$i^{t=1} = 2$}
( 5.8, 0) node [shape=rectangle,minimum size=0.5cm,draw] (obs3_1) {$i^{t=1} = 3$}
( -0.6,-1.3) node [shape=rectangle,minimum size=0.5cm,draw] (obs4_1) {$i^{t=1} = 4$}

( 1, 3) node [shape=tape,minimum size=0.5cm,draw,fill=lightgray] (obs1_2) {$i^{t=2} = 1$}
( -0.6,1.3) node [shape=rectangle,fill=lightgray,minimum size=0.5cm,draw] (obs1) {$i^{t=1} = 1$}
( -.3, 2.1) node [] {$p + \frac{(1-p)\times 2}{4+\gamma}$}
( 3.5, 1.3) node [] {$\frac{(1-p)\times 2}{4+\gamma}$}
( 6, 1.3) node [] {$\frac{(1-p)\times \gamma}{4+\gamma}$};

\draw[->,very thick,color=black
] (obs1_2) -- (table1);
\draw[->,very thick,color=black
] (obs1_2) -- (table2);
\draw[->,very thick,color=black
] (obs1_2) -- (table3);

\end{tikzpicture}

\vspace*{24pt}

\begin{tikzpicture}
\path
( 0, 0) node [shape=circle,minimum size=2cm,draw] (table1) {$k = 1$}
( 4, 0) node [shape=circle,minimum size=2cm,draw] (table2){$k = 2$}
( 8, 0) node [shape=circle,minimum size=2cm,draw] (table3){$k = 3$}
( 9.4, 0) node [shape=circle,draw,fill=black,minimum size =.01cm] {}
( 9.8, 0) node [shape=circle,draw,fill=black,minimum size =.01cm] {}
( 10.2, 0) node [shape=circle,draw,fill=black,minimum size =.01cm] {}

( -0.6,1.3) node [shape=rectangle,minimum size=0.5cm,draw] (obs1) {$i^{t=1} = 1$}
( 2.2, 0) node [shape=rectangle,minimum size=0.5cm,draw] (obs2_1){$i^{t=1} = 2$}
( 5.8, 0) node [shape=rectangle,minimum size=0.5cm,draw] (obs3_1) {$i^{t=1} = 3$}
( -0.6,-1.3) node [shape=rectangle,minimum size=0.5cm,draw] (obs4_1) {$i^{t=1} = 4$}

( 2.4, -1) node [shape=tape,minimum size=0.5cm,draw] (obs1_2_2) {$i^{t=2} = 1$}

( 1.2, 1.3) node [
] {$\frac{(1-p) \times 2}{5+\gamma}$}
( 4, 1.3) node [
] {$p +\frac{(1-p) \times 3}{5+\gamma}$}
( 7, 1.8) node [
] {$\frac{(1-p) \times \gamma}{5+\gamma}$}

( 1, 3) node [shape=tape,minimum size=0.5cm,draw,fill=lightgray
] (obs2_2) {$i^{t=2} = 2$}
( 2.2, 0) node [shape=rectangle,minimum size=0.5cm,draw,fill=lightgray
] (obs2_1){$i^{t=1} = 2$};

\draw[->,very thick,color=black
] (obs2_2) -- (table2);
\draw[->,very thick,color=black
] (obs2_2) -- (table1);
\draw[->,very thick,color=black
] (obs2_2) -- (8,1);

\end{tikzpicture}
\caption{\textbf{Intergenerational Chinese Restaurant Process.}
This figure illustrates the generative process of the latent clusters at $t = 2$.
The fractions near arrows represent the probabilities of units being assigned to the corresponding clusters.
The top and bottom panels illustrate the assignment process for unit 1 and unit 2  at $t=2$, respectively.
For illustration purposes, we assume that in the previous generation of $t = 1$, there are two units in cluster 1 ($i^{t=1} = 1$ and $i^{t=1} = 4$) and two units in cluster 2 ($i^{t=1} = 2$ and $i^{t=1} = 3$). In the bottom panel, we assume that unit 1 at $t=2$ goes to cluster 2. The stickiness between the two generations of $t=1$ and $t=2$ is defined by the parameter $p$.
}
\label{fig:igcrp}
\end{figure}

The dynamic processes at $t = 2, 3, ..., T$, defined by equations ~\eqref{eq:dynstkbrk_1} to \eqref{eq:dynstkbrk_6}, can be represented as a Chinese restaurant process with added stickiness, which is defined by the stickiness parameter $p$ . Figure~\ref{fig:igcrp} illustrates the process at $t = 2$.

The top panel of Figure~\ref{fig:igcrp} illustrates the cluster assignment process of unit 1($i^{t=2} = 1$), the first unit at $t = 2$.
For illustration purposes, we assume that in the previous generation of $t = 1$, there are two units in cluster 1 ($i^{t=1} = 1$ and $i^{t=1} = 4$) and two units in cluster 2 ($i^{t=1} = 2$ and $i^{t=1} = 3$). The stickiness between the two generations of $t=1$ and $t=2$ is defined by the parameter $p$. With probability $p$, unit 1 at $t = 2$ is assigned directly to cluster 1, where the previous generation of unit 1 at $t = 1$ stays; with probability $1 - p$, unit 1 does not go directly to cluster 1 but  instead receives a cluster assignment through a Chinese restaurant process, in which the number of units in an existing cluster is determined  by both the current generation of $t =2$ and the previous generation of $t =1$. If unit 1 does not go directly to cluster 1, it can still be assigned to cluster 1 with probability $2 / (4 + \gamma)$. Therefore, the total probability that unit 1 at $t = 1$ is assigned to cluster 1 is $p + (1 - p) \times (2 / (4 + \gamma))$. The probability that unit 1 is assigned to cluster 2 equals to ($1-p$) multiplies the conditional probability decided by the Chinese restaurant process, $2 / (4 + \gamma)$. Similarly, the probability that unit 1 is assigned to a new cluster is $(1 - p) \times 2 / (4 + \gamma)$. 

The bottom panel of Figure~\ref{fig:igcrp} illustrates the cluster assignment of unit $2$ at $t = 2$ ($i^{t=2} = 2$), assuming that unit 1 at $t = 1$ is assigned to cluster 2.
Again, unit $2$ may directly go to cluster 2, which unit 2 at $t = 1$ stays, or goes through a Chinese restaurant process. Altogether, the probability that unit 2 goes to cluster 2 is $p + (1-p) \times 3 / (5 + \gamma)$. The probability of unit 2 going to cluster 1 is $(1-p) \times 2 / (5 + \gamma)$, and the probability of unit 2 going to a new cluster is $(1-p) \times \gamma / (5 + \gamma)$. 

\subsection{Related Models}
Among models introducing time dynamics to DP mixture models, the proposed dynamic DP mixture model in our paper is most closely related to \citet{caron2012generalizedpolyaurntimevarying}. Both models introduce a stickiness parameter. 
However, in our model, cluster assignments at $t$ depend on the assignments of all units at $t-1$, while in \citet{caron2012generalizedpolyaurntimevarying}, cluster assignments at $t$ are determined by a selected subset of units at $t-1$, as influenced by the stickiness parameter. 
This difference results in stronger temporal dependence between generations in our model. For example, in the case of $p=0$ (i.e., when the cluster assignment of unit $i$ at $t$ is not directly influenced by its assignment at $t =1$), the cluster assignment process at $t$ is still influenced by cluster assignments of all units at $t - 1$. In \citet{caron2012generalizedpolyaurntimevarying}, no such influence exists in this case.

\cite{10.1214/22-AOAS1717} develops another dynamic DP model, in which $\pi$'s in Equation~\eqref{eq:dynstkbrk_q} follow an autoregressive process. In their model, cluster assignments do not depend on the assignments from the previous period.
Additionally, the cluster-specific parameters are not shared across time.
In \citet{huang2015dynamic}, units remain in the same cluster and exit from the dataset after a certain length of time.

The time-sensitive Dirichlet process mixture model developed by \citet{zhu2005time} does not have a stickiness parameter.

In some other models, the DP in period $t$ is a mixture of the DP at $t - 1$ and an entirely new DP \citep{10.1093/biostatistics/kxj025,doi:10.1198/jasa.2009.ap08497,10.1145/1835804.1835940,doi:10.1080/01621459.2021.1886105}. In other words, units at $t$ go into the same Chinese restaurant process as $t - 1$ or a brand-new Chinese restaurant process.

\subsection{Markov Chain Monte Carlo Algorithm for Estimation}
\label{mcmc}
In this section, we introduce the Markov Chain Monte Carlo (MCMC) algorithm used to estimate our model constructed through the dynamic stick-breaking process. Specifically, we employ the blocked Gibbs sampling algorithm, which consists of two major steps.

One step is to sample the cluster-specific parameter $\Theta_g$, which defines the data generating function $f(\Theta_k)$ for $k=1, 2, ...$, conditioned on the current cluster assignment $g[it] = k$.
Sampling $\Theta_k$ from its posterior distribution is straightforward when the prior distribution for $\Theta_k$ is conjugate.

The second step is to sample cluster assignments $g[it]$ for $i=1,2,...,N$ and $t=1,2,...,T$, conditioned on $\Theta_k$ for $k=1, 2, ...$.
Sampling $g[it]$ is more complex, as the posterior distribution is conditioned on both cross-sectional and temporal relationships of unit $i$ at time $t$ with other units. To address this complexity, We combine the forward-backward algorithm designed for change-point models \citep{chib98:_chang-poing_models} with the truncation approximation approach developed for stick-breaking priors \citep{Ishwara_James_01} to sample $g[it]$.

Following the truncation approximation approach of \citet{Ishwara_James_01}, we begin the algorithm by setting an arbitrarily large number $K$ at which we truncate the theoretically infinite number of clusters. In the posterior distributions of $g[it]$ for all $i = 1, 2, ..., N$ and $t = 1, 2, ..., T$, the probabilities for most clusters will be zero. This approach allows the model to estimate the number of clusters automatically.
With $K$ set, we initialise the starting values of $g[it]$ for all $i = 1, ..., N$ and $t = 1, ..., T$.

Each iteration of the Gibbs sampler then proceeds as follows.
The first step is to update $\Theta_{k}$ for $k = 1, 2, ..., K$. The posterior distribution of $\Theta_{k}$ is only conditioned on observed data $Y_{it}$ (and $X_{it}$ in the linear regression model example) for units in cluster $k$. The specific sampling algorithm for $\Theta_{k}$ depends on the form of function $f(\Theta_{g[it]})$.

Second, we sample the stickiness parameter $p$, which represents the probability that a unit goes directly to the same cluster as in the previous period. To simplify the sampling of $p$, we introduce augmented binary variables $d_{it}$ for $i = 1, 2, ..., N$ and $t = 1, 2, 3, ...T$.
These variables indicate whether a specific unit is directly assigned to the same cluster as before. The posterior distribution of $d_{it}$ is only conditioned on $g[it]$, $g[i,t-1]$, and $p$. Specifically, When $g[it] \neq g[i,t-1]$, $d_{it}$ must be 0; when $g[it] = g[i,t-1]$, unit $i$ may be assigned directly to the same cluster as before or assigned through a regular DP. Formally, 
\[ p(d_{it} = 1) =  \left\{ \begin{array}{ll}
0 & \mbox{if $g[it] \neq g[i,t-1]$};\\
\frac{p}{p + (1-p)q^t_k} & \mbox{if $g[it] = g[i,t-1]=k$}.\end{array} \right. \]
where $q^t_k$ is sampled in the previous MCMC iteration (described below).

Since we set the prior distribution for $p$ as a Beta distribution, the posterior distribution of $p$, conditioned on $d_{it}$ for all $i = 1, 2, ..., N$ and $t = 1, 2, ..., T$, is also a Beta distribution. Specifically, we sample $p$ as follows:
$$
p \sim \mathsf{Beta}(\alpha_p + N_1, \beta_p + N_2)
$$
$$
N_1 = \sum_{i=1}^N\sum_{t = 2}^T{d_{it}}
$$
$$
N_2 = \sum_{i=1}^N\sum_{t = 2}^T{(1-d_{it})}
$$

Third, we update the stick-breaking weights  $\pi_{k}^t$ and $q_k^t$ for $k = 1,2,...K$ and $t = 1, 2, ..., T$. Since $K$ is set to be an arbitrarily large number to approximate the stick-breaking prior with an infinite number of clusters, the posterior distribution of $\pi_{k}^t$ becomes a Beta distribution:
$$
\pi_k^t \sim \mathsf{Beta}(1+n_k^{t-1} + n_k^t, \gamma + \sum_{l = k+1}^K n_l^{t-1}+\sum_{l = k+1}^K n_l^t)
$$
$$
n_k^{t-1} = \sum_{i=1}^N{\mathbb{I}(g[i, t-1]=k)}
$$
$$
n_k^t = \sum_{i=1}^N{(1-d_{it})}\mathbb{I}(g[it]=k)
$$

Once $\pi_{k}^t$ is updated, calculate $q_k^t$:
$$
q_k^t = \pi_k^t \prod_{l = 1}^{k-1}(1-\pi_l^t)
$$

The last step is to sample cluster assignments $g[it]$ for $i = 1, 2, ..., N$ and $t = 1, 2, ..., T$.
Let $\mathbf{g}[t]\equiv (g[1t], g[2t], ..., g[Nt])'$, $\mathbf{q}^t\equiv (q_1^t, q_2^t, ..., q_K^t)'$, $\mathbf{Y}_t\equiv (Y_{1t}, Y_{2t}, ..., Y_{Nt})'$. Following \citet{chib98:_chang-poing_models}, we sample $\mathbf{g}[T]$,$\mathbf{g}[T-2]$, ...,$\mathbf{g}[1]$ in turn: 
\begin{align}
& Pr(g[it] = k | \mathbf{g}[T],...,\mathbf{g}[t+1], p,\mathbf{q}^T,..., \mathbf{q}^1 ,\mathbf{Y}_T, ... , \mathbf{Y}_1, \Theta_1, ..., \Theta_K) 
 \label{eq:backsamp_posterior} \\
 \propto &
\underset{\text{part } (1)}{\underbrace{ Pr(g[i,t+1]|p,g[it] = k, \mathbf{q}^{t+1})}}
\underset{\text{part } (2)}{\underbrace{Pr(g[it]=k|p, \mathbf{q}^{t},..., \mathbf{q}^1 ,\mathbf{Y}_t, ... , \mathbf{Y}_1,\Theta_1, ..., \Theta_K)}}\label{eq:forfilter_1} 
\end{align}

The equation above shows the decomposition of the conditional posterior distribution of $g[it]$ given the observed data, model parameters, and cluster assignments for period $t + 1$ through $T$. Part (1) represents the probability of $g[i,t+1]$ given that $g[i,t]$ is in cluster $k$. Specifically,
$$
Pr(g[i,t+1] = l|p,g[it] = k, q_l^{t+1}) = (1-p)q^{t+1}_l + p\mathbb{I}(l=k)
$$

Part (2) represents the conditional probability of $g[it]=k$ given the model parameters and the data from period 1 up to period $t$. Part 2 can be further decomposed as:
$$
Pr(g[it]=k|p, \mathbf{q}^{t},..., \mathbf{q}^1 ,\mathbf{Y}_t, ... , \mathbf{Y}_1,\Theta_1, ..., \Theta_K))
$$
$$ \propto \underset{\text{part } a}{\underbrace{ f(Y_{it}| g[it]=k, \mathbf{Y}_{t-1}, ... , \mathbf{Y}_1, \Theta_1, ..., \Theta_K)}}
\underset{\text{part } b}{\underbrace{Pr(g[it]=k|p, \mathbf{q}^{t},..., \mathbf{q}^1 ,\mathbf{Y}_{t-1}, ... , \mathbf{Y}_1,\Theta_1, ..., \Theta_K)}} 
$$

Part (a) is the distribution of observed data $Y_{it}$ conditioned on the cluster assignment $g[it]$, observed data from all the previous periods, and cluster-specific parameters $\Theta_k$ for $k = 1, 2, ..., K$. Given $g[it]=k$ and $\Theta_k$, the distribution of $Y_{it}$ is conditionally independent of observed data from the previous periods:
$$
f(Y_{it}| g[it]=k, \mathbf{Y}_{t-1}, ... , \mathbf{Y}_1,\Theta_1, ..., \Theta_K) =  f(Y_{it}|\Theta_k)
$$

Compared to the conditional distribution of $g[it]$ in Part (2), in Part (b), $g[it]$ is no longer conditioned on $\mathbf{Y}_{t}$. Part (b) can be further decomposed as:
$$
Pr(g[it]=k|p,\mathbf{q}^{t},..., \mathbf{q}^1 ,\mathbf{Y}_{t-1}, ... , \mathbf{Y}_1,\Theta_1, ..., \Theta_K)
$$
$$
= \sum_{l} \underset{\text{part } (c)} { \underbrace{Pr(g[it]=k|p,q^t_k, g[i,t-1]=l)}}  \underset{\text{part } (d)} {\underbrace{Pr(g[i,t-1]=l|p,\mathbf{q}^{t-1},..., \mathbf{q}^1 ,\mathbf{Y}_{t-1}, ... , \mathbf{Y}_1,\Theta_1, ..., \Theta_K)}}
$$

There is a recursive structure between the equation above and part (1) and part (2) in equation~\eqref{eq:forfilter_1}.
Therefore, the standard forward recursion algorithm computes the conditional posterior given by equation~\eqref{eq:backsamp_posterior}, while the backward sampling algorithm generates MCMC draws from the conditional posterior of the cluster assignments.

\section{Empirical Application}
\label{sec:analysis}
In this section, we apply our proposed dynamic DP mixture model to analyse racial realignment in the twentieth century, using data on legislative activities from \citet{Schickle16}. We begin by explaining how our model is tailored to fit the specific data used in the racial realignment case. Next, we describe the details of the MCMC algorithm employed for estimation. Finally, we present the results of our analysis.

\subsection{Statistical Model for Analyzing Racial Realignment}
The aim of applying our proposed model to the case of racial realignment is to analyse how state parties allied with each other. We assume that a state party's position on civil rights issues is reflected in its Congress members' legislative activities. Although some Congress members may engage in activities that go against their party's general consensus at the state level, we regard these situations as idiosyncratic. Fundamentally, we assume that Congress members' legislative activities are largely constrained by their state parties. Based on this assumption, we can infer the latent clustering patterns of state parties and how these patterns changed over time using data on legislative activities in the House of Representatives.

Let $t = 73, 74, ..., 92$ index the 73rd Congress (1933-35) to the 92nd Congress (1971-73) and let $s = 1, 2, ..., 50$ index the 50 states. Let $D$ and $R$ represent the Democratic Party and the Republican Party, respectively. Let $i = 1, 2, ..., I^D_{st} (\text{or } I^R_{st}) $ represent a Democratic (or Republican) Congress member from state $s$ in the $t$-th Congress, where $I^D_{st}$ is the total number of Democrats in the House of Representatives from State $s$ in the $t$-th Congress, and  $I^R_{st}$ represents the total number of Republicans correspondingly.  Let $j = 1, 2, ..., J_t^V (\text{or } J_t^P$) index a civil rights vote call or petition in the $t$-th Congress, where $J_t^V$ represents the total number of civil rights vote calls, and $J_t^P$ the total number of petitions correspondingly. 

Let $g^D[st]$ represent the cluster which the Democratic Party of state $s$ in the $t$-th Congress belongs. Similarly, let $g^R[st]$ represent the cluster of the Republican party. We assume $g^D[st]$ and $g^R[st]$ share the same dynamic DP prior as as defined in Section \ref{sec:model}. Here we explain the specific statistical model required in Equation ~\eqref{eq:stkbrk_1} for modelling observed data on legislative activities.

\textbf{Roll-call Votes.} Let $V^D_{istj}$ represent roll-call vote $j$ in the $t$-th Congress for Democratic member $i$ from state $s$. Similarly, let $V^R_{istj}$ represent the roll-call vote for a Republican member. We assume $V^D_{istj}$  and $V^R_{istj}$ follow the following distributions:
$$
V^D_{istj} \overset{ind.}{\sim} \mathsf{Bernoulli} (\theta_{g^D[st]})
$$
$$
V^R_{istj} \overset{ind.}{\sim} \mathsf{Bernoulli} (\theta_{g^R[st]})
$$
For $g^D[st]=1, 2,...,k...$ and $g^R[st]=1, 2,...,k...$,
$$
\theta_k \sim \mathsf{Beta}(\alpha_{\theta}, \beta_{\theta})
$$

\textbf{Discharge Petitions} Let $P^D_{istj}$ and $P^R_{istj}$ represent whether member $i$ from state $s$ signed petition $j$ in the $t$-th Congress, where $D$ and $R$ index Democrats and Republicans, respectively. We assume that $P^D_{istj}$ and 
$P^R_{istj}$ follow the following distributions:
$$
P^D_{istj} \overset{ind.}{\sim} \mathsf{Bernoulli} (\eta_{g^D[st]})
$$
$$
P^R_{istj} \overset{ind.}{\sim} \mathsf{Bernoulli} (\eta_{g^R[st]})
$$
For $g^D[st]=1, 2,...,k...$ and $g^R[st]=1, 2,...,k...$,
$$
\eta_k \sim \mathsf{Beta}(\alpha_{\eta}, \beta_{\eta})
$$

\textbf{Floor Speeches.} Let $S^D_{ist}$ and $S^R_{ist}$ represent whether Democratic or Republican member $i$ from state $s$ delivered at least one floor speech in support of civil rights during the $t$-the Congress. We assume that $S^D_{ist}$ and $S^R_{ist}$ follow the following distributions: 
$$
S^D_{ist} \overset{ind.}{\sim} \mathsf{Bernoulli} (\omega_{g^D[st]})
$$
$$
S^R_{ist} \overset{ind.}{\sim} \mathsf{Bernoulli} (\omega_{g^R[st]})
$$
For $g^D[st]=1, 2,...,k...$ and $g^R[st]=1, 2,...,k...$,
$$
\omega_k\sim \mathsf{Beta}(\alpha_{\omega}, \beta_{\omega})
$$

\textbf{Bill Sponsorship.} Finally, let $B^D_{ist}$ and $B^R_{ist}$ represent the number of civil rights bills sponsored by Democratic or Republican member $i$ from state $s$ in the $t$-th Congress. We assume $B^D_{ist}$ and $B^R_{ist}$ follow the following distributions:

$$
B^D_{ist} \overset{ind.}{\sim} \mathsf{Poisson} (\lambda_{g^D[st]})
$$
$$
B^R_{ist} \overset{ind.}{\sim} \mathsf{Poisson} (\lambda_{g^R[st]})
$$
For $g^D[st]=1, 2,...,k...$ and $g^R[st]=1, 2,...,k...$,
$$
\lambda_k\sim \mathsf{Gamma}(a_{\lambda}, b_{\lambda})
$$

\subsection{MCMC Algorithm}
As introduced in Section \ref{mcmc}, the MCMC algorithm includes two major steps. One step is to sample the cluster-specific parameters, here $\theta_k$, $\eta_k$, $\omega_k$, and $\lambda_k$, for $k = 1, 2, ..., $ conditioned on cluster assignments $g^D[st]$ and $g^R[st]$ for $s = 1, 2, ..., 50$ and $t = 73, 74, ..., 92$. We set conjugate priors for the cluster-specific parameters. Given the cluster assignments, the posterior distributions of $\theta_k$, $\eta_k$, $\omega_k$ are Beta distributions, and the posterior distribution of $\lambda_k$ is a Gamma distribution. The second step is to sample the cluster assignments $g^D[st]$ and $g^R[st]$ for $s = 1, 2, ..., 50$ and $t = 73, 74, ..., 92$, conditioned on the cluster-specific parameters. To achieve this, We combine the forward-backward approach for change-point models and the truncation approximation approach for stick-breaking priors. Details of the algorithm are shown in SI~\ref{mcmc2}.

\subsection{Empirical Results}
The MCMC outputs of our model provide the cluster assignments of all state parties in each congressional session. With this information, we can analyse racial alignment patterns across the North-South division, along party lines, or both, and examine how these patterns change over time.

We first investigate whether our model replicates the findings of \citet{Schickle16}, which suggest that in the North, the Democratic Party and the Republican Party began to diverge in their civil rights positions during the 1940s. To achieve this, we analyse whether and when, in the North, the Democratic Party and the Republican Party at the state level were less likely to belong to the same cluster. We perform a similar analysis for the South to examine the pattern of cross-party alignment and how it changes over time.

Next, we test two arguments frequently mentioned by \citet{Schickle16} but not directly tested with legislative data. The first argument is that the Democratic Party's North-South coalition dissolved long before the 1960s. The second argument is that the Republican Party in the North became increasingly divided over civil rights issues before the 1960s. We test these arguments by analyzing the following: (1) Within the Democratic Party, whether and when northern state parties were less likely to be in the same cluster as southern state parties; (2) Within the Northern Republican Party, whether and when northern state parties were less likely to be in the same cluster.

Finally, we examine the solidarity of the Democratic Party in the South—an assumption implicitly made by \citet{Schickle16} but not tested—by analyzing how the probability of two Southern states being in the same cluster within the Democratic Party changes over time.

\begin{figure}[tp!]
    \centering
    \caption{Cross-party Alignment, in the North and South}

    \begin{subfigure}[b]{0.8\textwidth}
        \centering
        \includegraphics[width=\textwidth]{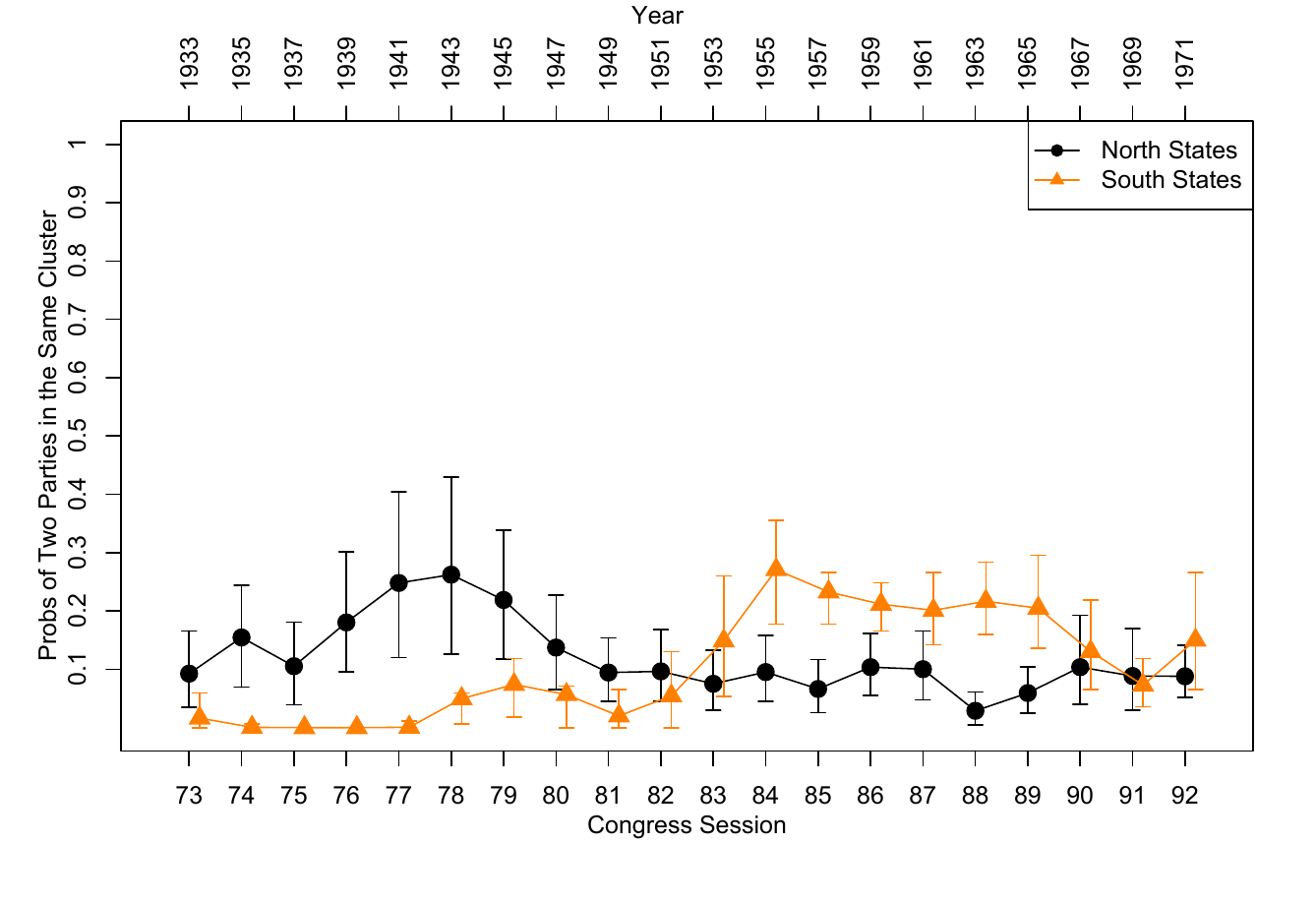} 
        \caption*{(a) \textbf{Pooled Results.} Each point represents the average probability that a state Democratic Party and a state Republican Party are in the same cluster in the North (or South) for the corresponding Congress, with a 95\% credible interval.}
        \vspace{0.5em}
    \end{subfigure}
    \vspace{1.5em} 
    \begin{subfigure}[b]{0.8\textwidth}
        \centering
        \includegraphics[width=\textwidth]{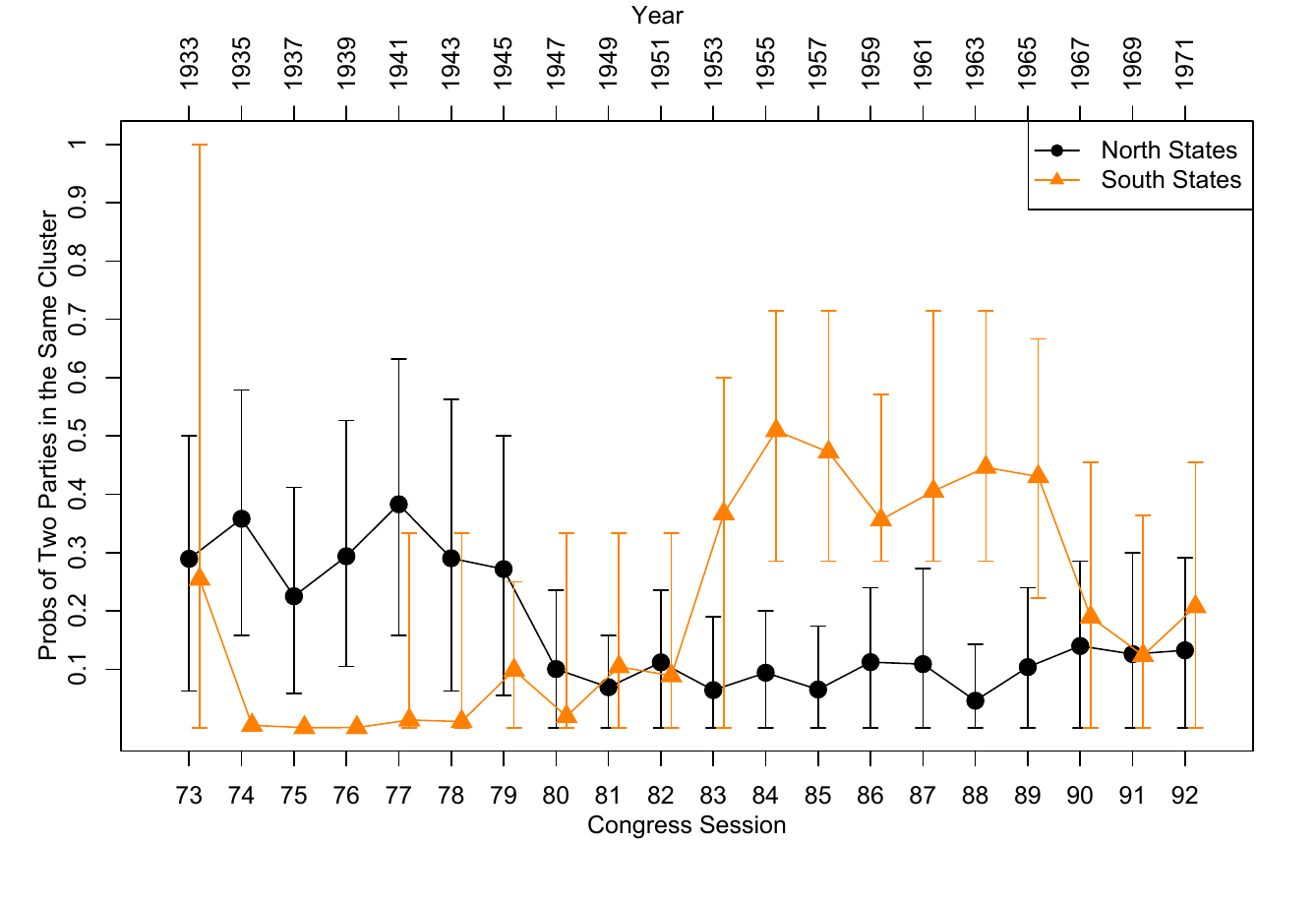} 
        \caption*{(b) \textbf{Within-state Results.} Each point represents the average probability that, within a given state, the Democratic Party and the Republican Party are in the same cluster for the corresponding Congress, with a 95\% credible interval.}
        \vspace{0.5em}
    \end{subfigure}
\label{fig:demEqualRep}
\end{figure}

\textbf{Cross-party Alignment.} The patterns of cross-party alignment in the North and South respectively are presented in Figure \ref{fig:demEqualRep}. Panel (a) shows the pooled results: each point represents the average probability that a state Democratic Party and a state Republican Party are in the same cluster in the North (or South) for the corresponding Congress, with a 95\% credible interval. Panel (b) shows the within-state results: each point represents the average probability that, within a given state, the Democratic Party and the Republican Party are in the same cluster for the corresponding Congress, with a 95\% credible interval. 

The pooled results and within-state results show consistent patterns of cross-party alignments. First, the probabilities that the two parties in the North are in the same cluster decline rapidly in the 1940s. This replicates the findings in \citet{Schickle16}, which indicate that by the end of the 1940s, the two parties in the North had diverged in their positions on civil rights.

Panel (a) shows that toward the end of the 1930s, northern Democrats are more likely to be in the same cluster as northern Republicans. This aligns with Schickler's finding that after the New Deal, northern Democrats quickly caught up with Republicans in supporting civil rights. 

Panel (b) shows that Democratic and Republican parties within the same Northern state have already had a high probability of acting together even before the New Deal. 

In addition to analyzing cross-party alignment in the North, as discussed in \citet{Schickle16}, we also perform the same analysis for Southern states. Interestingly, both pooled and within-state results show that in early 1950s, the probabilities that the two parties in the South are in the same cluster increased rapidly. Previously, the probabilities were very low (close to 0).

This result contradicts Schickler's claim that Southern Republicans began aligning with racially conservative Southern Democrats only after conservatives within the GOP captured the southern party machinery established by President Dwight Eisenhower (Chapter 10). Our results show that, even before Eisenhower's party-building efforts, Southern Republicans had already aligned with Southern Democrats.

Finally, while panel (a) explores the general pattern of cross-party alignment over civil rights issues in the North and South, panel (b) examines the within-state alignment pattern.  This allows us to investigate whether, when facing the same socioeconomic conditions, Democrats and Republicans are more likely to align with each other on civil rights issues. In general, the probabilities of the two parties in the same cluster are higher in panel (b) than in panel (a), indicating that within the same state, the two parties are more likely to take similar positions on civil rights issues.

\begin{figure}[t!]
\centering
\includegraphics[width=0.8\textwidth]{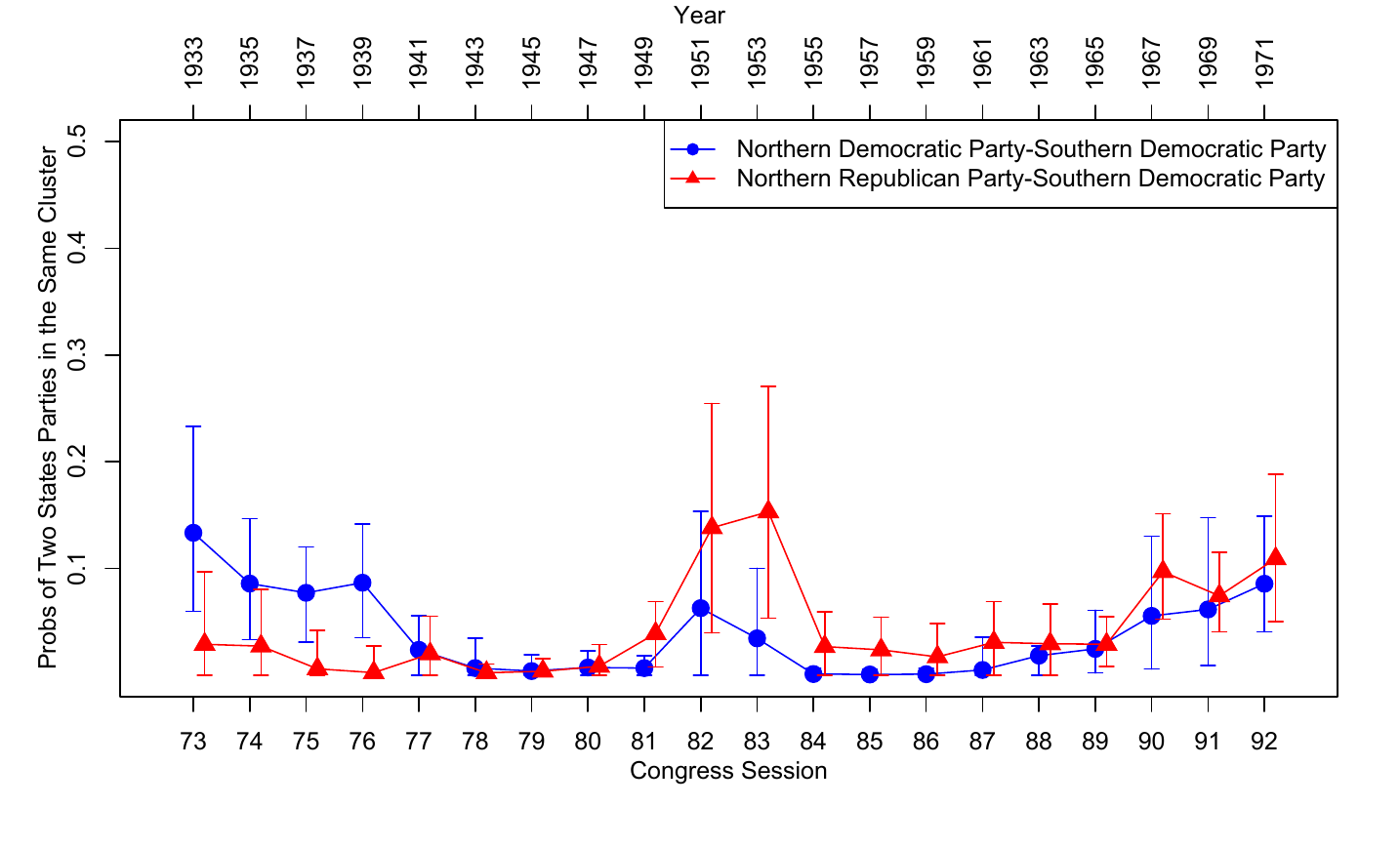}
\caption{\textbf{Alignment of Southern Democrats with Northern Democrats and Republicans Respectively.} Each point represents the average probability that a state Democratic Party (or Republican Party) in the North and a state Democratic Party in the South are in the same cluster, with a 95\% credible interval.
}
\label{fig:equSouth}
\end{figure}

\textbf{Alignment with Southern Democrats.} We now turn to evaluate an argument frequently made by \citet{Schickle16}, but not tested directly, that the Democratic Party's North-South coalition declined long before the 1960s. In Figure~\ref{fig:equSouth}, we calculate the average probability that a northern Democratic Party is in the same cluster as a southern Democratic Party in each year. For comparison, we also calculate the average probability that a northern Republican Party is in the same cluster as a southern Democratic Party, since Schickler argues that the GOP in some northern states, particularly in the rural Midwest, had already been more likely to align with southern Democrats before the 1960s.

Figure~\ref{fig:equSouth} shows that, consistent with Schickler's argument, the probability that northern and southern Democrats are in the same cluster declines rapidly in the 1930s. By the early 1940s, northern Democrats, like northern Republicans, have almost no chance of being in the same cluster as southern Democrats.

Toward the late 1940s and early 1950s, both northern Democrats and northern Republicans suddenly became more likely to align with southern Democrats. However, compared to northern Democrats, northern Republicans had a higher probability of being in the same cluster as southern Democrats during this period of peak alignment. After this period of sudden change, the probabilities of northern Republicans and southern Democrats being in the same cluster remain higher than those of northern and southern Democrats being in the same cluster.

\begin{figure}[t!]
\centering
\includegraphics[width=0.8\textwidth]{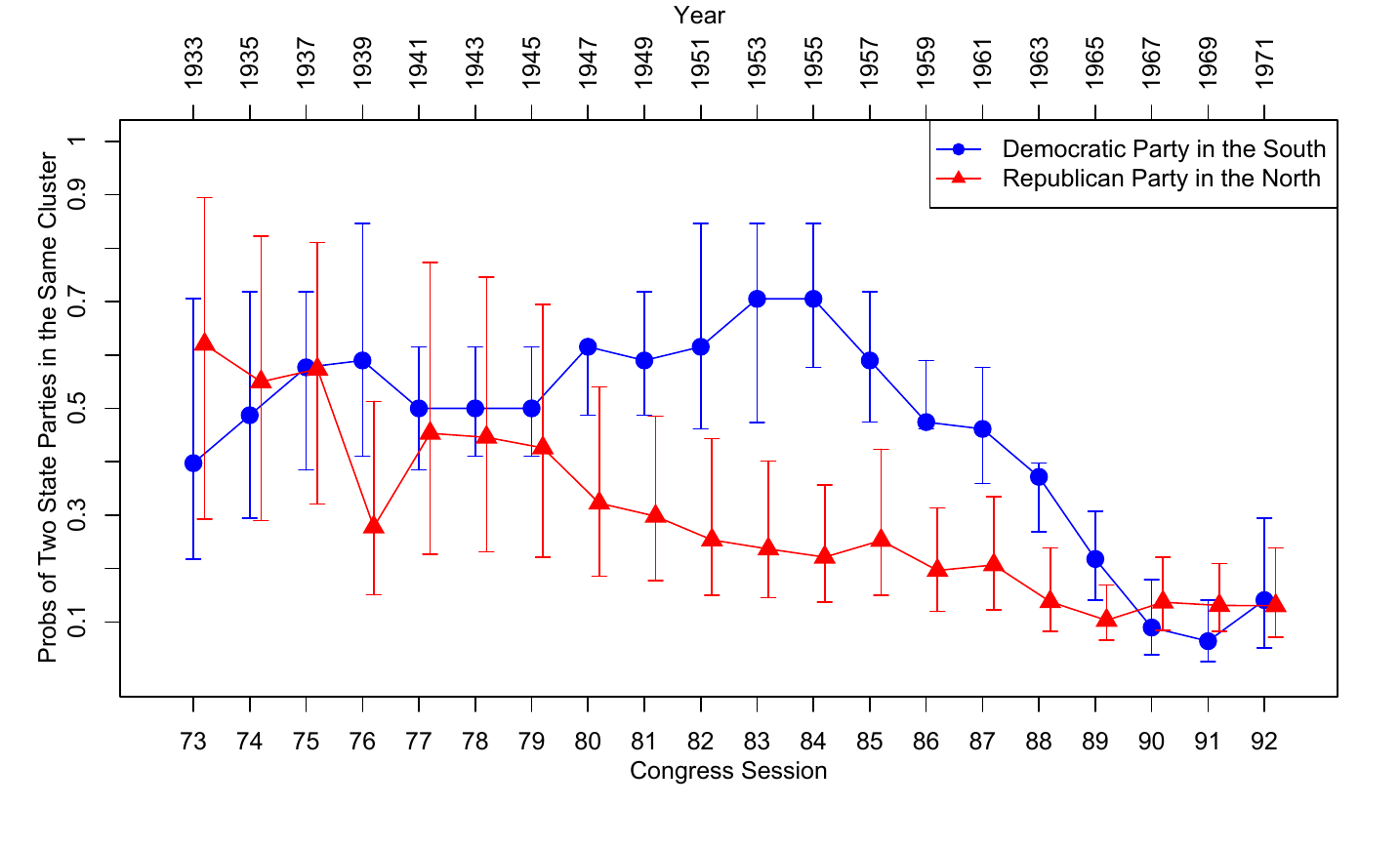}
\caption{\textbf{Within-party Solidarity, Northern Republicans and Southern Democrats.} Each point represents the average probability that two Republican parties in the North (or Democratic parties in the South) are in the same cluster, with a 95\% credible interval.
}
\label{fig:solidarity}
\end{figure}

\textbf{Within-Party Solidarity: Northern Republicans and Southern Democrats.} Finally, we examine another untested argument in Schickler (2016) that Northern Republicans had already become divided in the 1940s and 1950s over civil rights issues. Additionally, we evaluate the solidarity of Southern Democrats—an assumption made by Schickler (2016) but not formally tested. In Figure~\ref{fig:solidarity}, we calculate the average probability that a northern Republican Party in one state is in the same cluster as the Republican Party in another northern state in each year. Similarly, we calculate the average probability that a southern Democratic Party in one state is in the same cluster as the Democratic Party in another southern state in each year.

Consistent with Schickler (2016), Figure~\ref{fig:solidarity} shows that the probability of two northern Republican parties being in the same cluster declines steadily in the 1940s and 1950s. However, in contrast to Schickler's assumption that Southern Democrats were a solid unity against civil rights, their regional coalition peaked in the early 1950s and began to decline rapidly after 1955, during the tumultuous years of the civil rights movement.

\section{Concluding Remarks}
Most societal changes are evolutionary rather than revolutionary, and it is difficult to capture gradual changes as opposed to abrupt ones.
Debate over the timing of the racial realignment between two major parties in the United States illustrates this difficulty.
Although there was a clearly visible change at the level of party leadership in the 1960s, on which the conventional wisdom is based, statistical analysis of newly available data suggests that the realignment was a gradual process that began in the 1930s.

This paper develops a novel dynamic Dirichlet mixture model for analyzing evolutionary changes.
In the proposed model, units are clustered into latent clusters, and cluster assignments are assumed to follow a Markov process.
In each time period, a unit remains in the same cluster as the previous period with a probability controlled by a stickiness parameter or moves to another cluster with a probability determined by the Chinese restaurant process, conditional on the cluster assignments in the previous period. Because of this cluster assignment process, the proposed dynamic DP model can effectively capture gradual structural changes.

Our empirical analysis of racial realignments demonstrates that the switch in positions on civil rights between the Democratic Party and the Republican Party had been a gradual process since the New Deal, rather than a sudden structural break in the 1960s.
We found that the Democratic Party and the Republican Party began to take divergent positions on civil rights issues in the 1940s.
Moreover, our analysis identifies two key moments of racial realignment that have not been tested with data in existing studies.
First, at the end of the 1940s and the beginning of the 1950s, both Northern and Southern Republicans were more likely to be aligned with Southern Democrats.
President Truman's postwar civil rights advocacy during the 1948 election, which triggered the Dixiecrat bolt and deep Southern anger toward the Truman administration, may have facilitated this coalition between Republicans and Southern Democrats.
Second, during the Civil Rights Movement, starting in 1954, solidarity within the Southern Democratic Party gradually dissolved.

Beyond the case of racial realignment, the proposed model is widely applicable to any dataset with repeated measurements of multiple units.
Its use would therefore benefit many social scientists interested in the temporal dynamics of unobserved heterogeneity.

\newpage
\pdfbookmark[1]{References}{References}
\bibliography{dphmmix,rf_xiang,methods}

\clearpage
\onehalfspacing

\appendix
\counterwithin{figure}{section}
\counterwithin{table}{section}

\section{Summary Statistics}
\label{tab:datasummary}

\begingroup
\setlength{\tabcolsep}{6 pt} 
\renewcommand{\arraystretch}{0.5} 
\setstretch{0.5}

\begin{longtable}{|m{0.7cm}|m{1.5cm}|m{0.5cm}m{1.2cm}m{1.2cm}|m{0.5cm}m{1.2cm}m{1.2cm}|m{1.2cm}m{1.2cm}|m{1.2cm}m{1.2cm}|}
\caption{Summary Statistics of Four Types of Legislative Activities Related to Civil Rights in the U.S. House of Representatives (1933–1973)} \\
\hline

Con. & Year & \multicolumn{3}{|c|}{Roll Calls} & \multicolumn{3}{|c|}{Petitions}   & \multicolumn{2}{|c|}{Speeches} & \multicolumn{2}{|c|}{Bill Sponsorship}  \\ 
& & No.& Dem. & Rep. & No. & Dem. & Rep. & Dem. & Rep. & Dem. & Rep.\\
\hline
73 & 1933-35 & 0 &  &  & 1 & 0.156 (0.363) & 0.717 (0.453) & 0.025 (0.156) & 0.025 (0.157) &  &  \\ 
74 & 1935-37 & 0 &  &  & 1 & 0.392 (0.489) & 0.771 (0.422) & 0.037 (0.189) & 0.029 (0.167) &  &  \\ 
75 & 1937-39 & 2 & 0.507 (0.5) & 0.521 (0.501) & 2 & 0.224 (0.417) & 0.618 (0.487) & 0.059 (0.236) & 0.097 (0.297) &  &  \\ 
76 & 1939-41 & 2 & 0.476 (0.5) & 0.964 (0.186) & 3 & 0.161 (0.368) & 0.354 (0.479) & 0.052 (0.222) & 0.096 (0.295) &  &  \\ 
77 & 1941-43 & 3 & 0.589 (0.492) & 0.975 (0.157) & 4 & 0.16 (0.366) & 0.222 (0.416) & 0.029 (0.169) & 0.018 (0.132) &  &  \\ 
78 & 1943-45 & 2 & 0.478 (0.5) & 0.924 (0.265) & 4 & 0.178 (0.383) & 0.209 (0.407) & 0.022 (0.147) & 0.005 (0.068) &  &  \\ 
79 & 1945-47 & 3 & 0.531 (0.499) & 0.895 (0.307) & 4 & 0.305 (0.46) & 0.312 (0.464) & 0.036 (0.188) & 0.015 (0.123) &  &  \\ 
80 & 1947-49 & 1 & 0.422 (0.495) & 0.941 (0.235) & 2 & 0.19 (0.393) & 0.081 (0.273) & 0.051 (0.22) & 0.004 (0.063) & 0.102 (0.606) & 0.087 (0.399) \\ 
81 & 1949-51 & 4 & 0.567 (0.496) & 0.83 (0.376) & 3 & 0.234 (0.424) & 0.076 (0.265) & 0.053 (0.224) & 0.023 (0.149) & 0.132 (0.768) & 0.08 (0.292) \\ 
82 & 1951-53 & 0 &  &  & 1 & 0.05 (0.218) & 0.019 (0.138) & 0.074 (0.263) & 0.024 (0.154) & 0.107 (0.536) & 0.053 (0.33) \\ 
83 & 1953-55 & 0 &  &  & 2 & 0.272 (0.445) & 0.045 (0.208) & 0.073 (0.261) & 0.041 (0.198) & 0.178 (0.79) & 0.032 (0.199) \\ 
84 & 1955-57 & 1 & 0.522 (0.501) & 0.876 (0.33) & 1 & 0.403 (0.491) & 0.232 (0.423) & 0.14 (0.348) & 0.039 (0.195) & 0.43 (1.625) & 0.074 (0.358) \\ 
85 & 1957-59 & 2 & 0.55 (0.498) & 0.896 (0.306) & 2 & 0.154 (0.361) & 0.071 (0.257) & 0.142 (0.349) & 0.078 (0.27) & 0.446 (1.494) & 0.118 (0.428) \\ 
86 & 1959-61 & 5 & 0.591 (0.492) & 0.87 (0.337) & 1 & 0.564 (0.497) & 0.289 (0.455) & 0.143 (0.351) & 0.075 (0.265) & 0.254 (1.022) & 0.176 (0.792) \\ 
87 & 1961-63 & 1 & 0.655 (0.476) & 0.855 (0.353) & 0 &  &  & 0.095 (0.294) & 0.023 (0.149) & 0.374 (1.453) & 0.158 (1.004) \\ 
88 & 1963-65 & 2 & 0.635 (0.482) & 0.861 (0.347) & 2 & 0.283 (0.451) & 0.066 (0.249) & 0.182 (0.386) & 0.143 (0.351) & 0.517 (1.642) & 0.522 (1.169) \\ 
89 & 1965-67 & 2 & 0.761 (0.427) & 0.793 (0.406) & 0 &  &  & 0.225 (0.418) & 0.161 (0.369) & 0.225 (0.684) & 0.629 (0.845) \\ 
90 & 1967-69 & 0 &  &  & 0 &  &  & 0.23 (0.422) & 0.053 (0.224) & 0.254 (0.953) & 0.026 (0.191) \\ 
91 & 1969-71 & 0 &  &  & 1 & 0.46 (0.499) & 0.11 (0.314) &  &  & 0.352 (0.951) & 0.14 (0.481) \\ 
92 & 1971-73 & 0 &  &  & 1 & 0.318 (0.467) & 0.406 (0.492) &  &  & 0.512 (1.813) & 0.299 (0.993) \\ 
\hline
\multicolumn{12}{p{18cm}}{Note: This table shows summary statistics of four types of legislative activities related to civil rights, by Congress and by political parties. Standard deviations are shown in parentheses. Variables of roll calls are dummies  indicating whether a member voted ``yes'' for a civil rights bill; variables of petitions are dummies for signing a discharge petition for advancing a civil right bill; variables of speeches are dummies indicating whether a member delivered at least one pro-civil rights speech during a certain Congress; bill sponsorship is a count variable measuring how many civil rights bills a member sponsored during a certain Congress.}
\label{tab:sum}
\end{longtable}

\endgroup

\clearpage

\clearpage
\section{Markov Chain Monte Carlo Algorithm for the Empirical Application}
\label{mcmc2}
In Section~\ref{mcmc}, we introduced the Markov Chain Monte Carlo (MCMC) algorithm for a general dynamic DP mixture model. In this section, we present a specific algorithm tailored to the empirical application. Most components of this algorithm are identical to those in the general algorithm. Here, we explain the parts that are specific to the application example.

As in Section \ref{mcmc}, at the beginning of the algorithm, set an arbitrarily large number $K$ to truncate the number of clusters. Then, initialize the starting values of $g^D[st]$ and $g^R[st]$ for $s = 1, 2, ..., 50$ and $t = 73, 74, ..., 92$. After initialization, each iteration of the Gibbs sampler proceeds as follows:
\begin{enumerate}
	\item Update $\theta_k, \eta_k, \omega_k, \lambda_k$ for $k=1, 2, ..., K$
	\begin{enumerate}
		\item The Posterior Distribution of $\theta_k$
		$$
		\theta_k \sim \mathsf{Beta}(\alpha_{\theta} + N_{\theta}^1,  \beta_{\theta} + N_{\theta}^0)
		$$
		$$N_{\theta}^1 = \sum_{t=73}^{92} \sum_{s=1}^{50} \sum_{i=1}^{I_{st}^D} \sum_{j=1}^{J_t^V} V_{istj}^D \mathbb{I}(g^D[st]=1) + \sum_{t=73}^{92} \sum_{s=1}^{50} \sum_{i=1}^{I_{st}^R} \sum_{j=1}^{J_t^V} V_{istj}^R \mathbb{I}(g^R[st]=1)
		$$
		$$N_{\theta}^0 = \sum_{t=73}^{92} \sum_{s=1}^{50} \sum_{i=1}^{I_{st}^D} \sum_{j=1}^{J_t^V} (1-V_{istj}^D) \mathbb{I}(g^D[st]=1) + \sum_{t=73}^{92} \sum_{s=1}^{50} \sum_{i=1}^{I_{st}^R} \sum_{j=1}^{J_t^V} (1-V_{istj}^R) \mathbb{I}(g^R[st]=1)
		$$
		\item The Posterior Distribution of $\eta_k$
		$$
		\eta_k \sim \mathsf{Beta}(\alpha_{\eta} + N_{\eta}^1,  \beta_{\eta} + N_{\eta}^0)
		$$
		$$N_{\eta}^1 = \sum_{t=73}^{92} \sum_{s=1}^{50} \sum_{i=1}^{I_{st}^D} \sum_{j=1}^{J_t^P} P_{istj}^D \mathbb{I}(g^D[st]=1) + \sum_{t=73}^{92} \sum_{s=1}^{50} \sum_{i=1}^{I_{st}^R} \sum_{j=1}^{J_t^P} P_{istj}^R \mathbb{I}(g^R[st]=1)
		$$
		$$N_{\theta}^0 = \sum_{t=73}^{92} \sum_{s=1}^{50} \sum_{i=1}^{I_{st}^D} \sum_{j=1}^{J_t^P} (1-P_{istj}^D) \mathbb{I}(g^D[st]=1) + \sum_{t=73}^{92} \sum_{s=1}^{50} \sum_{i=1}^{I_{st}^R} \sum_{j=1}^{J_t^P} (1-P_{istj}^R) \mathbb{I}(g^R[st]=1)
		$$
		\item The Posterior Distribution of $\omega_k$
		$$
		\omega_k \sim \mathsf{Beta}(\alpha_{\omega} + N_{\omega}^1,  \beta_{\omega} + N_{\omega}^0)
		$$
		$$N_{\omega}^1 = \sum_{t=73}^{92} \sum_{s=1}^{50} \sum_{i=1}^{I_{st}^D}S_{ist}^D \mathbb{I}(g^D[st]=1) + \sum_{t=73}^{92} \sum_{s=1}^{50} \sum_{i=1}^{I_{st}^R} S_{ist}^R \mathbb{I}(g^R[st]=1)
		$$
		$$N_{\omega}^0 = \sum_{t=73}^{92} \sum_{s=1}^{50} \sum_{i=1}^{I_{st}^D}(1-S_{ist}^D) \mathbb{I}(g^D[st]=1) + \sum_{t=73}^{92} \sum_{s=1}^{50} \sum_{i=1}^{I_{st}^R} (1-S_{ist}^R) \mathbb{I}(g^R[st]=1)
		$$
		\item The Posterior Distribution of $\lambda_k$
		$$
		\lambda_k \sim \mathsf{Gamma}(\alpha_{\omega} + C_{\lambda},  \beta_{\omega} + N_{\lambda})
		$$
		$$C_{\lambda} = \sum_{t=73}^{92} \sum_{s=1}^{50} \sum_{i=1}^{I_{st}^D}B_{ist}^D \mathbb{I}(g^D[st]=1) + \sum_{t=73}^{92} \sum_{s=1}^{50} \sum_{i=1}^{I_{st}^R} B_{ist}^R \mathbb{I}(g^R[st]=1)
		$$
		$$N_{\lambda} = \sum_{t=73}^{92} \sum_{s=1}^{50} \sum_{i=1}^{I_{st}^D} \mathbb{I}(g^D[st]=1) + \sum_{t=73}^{92} \sum_{s=1}^{50} \sum_{i=1}^{I_{st}^R}\mathbb{I}(g^R[st]=1)
		$$
	\end{enumerate}
	\item Sample the Stickiness Parameter $p$\\
	To sample $p$, we first introduce a series of dummy variables $d^D_{st}$ and $d^R_{st}$ for $t = 2, 3, ...T$ and $i = 1, 2, ..., N$ to indicate whether a unit directly stays in the same cluster as in the previous period. 
	\begin{enumerate} 
		\item Sample $d^D_{st}$ and $d^D_{st}$
		\[ p(d^D_{st} = 1) =  \left\{ \begin{array}{ll}
		0 & \mbox{if $g^D[st] \neq g^D[s,t-1]$};\\
		\frac{p}{p + (1-p)q^t_k} & \mbox{if $g^D[st] = g^D[s,t-1]=k$}.\end{array} \right. \] 
		\[ p(d^R_{st} = 1) =  \left\{ \begin{array}{ll}
		0 & \mbox{if $g^R[st] \neq g^R[s,t-1]$};\\
		\frac{p}{p + (1-p)q^t_k} & \mbox{if $g^R[st] = g^R[s,t-1]=k$}.\end{array} \right. \] 

		\item Sample p
		$$
		p \sim \mathsf{Beta}(\alpha_p + N_1, \beta_p + N_2)
		$$
		$$
		N_1 = \sum_{s=1}^{50}\sum_{t = 72}^{93}{d^D_{st}} + \sum_{s=1}^{50}\sum_{t = 72}^{93}{d^R_{st}}
		$$
		$$
		N_2 = \sum_{s=1}^{50}\sum_{t = 72}^{93}{(1-d^D_{st})} + \sum_{s=1}^{50}\sum_{t = 72}^{93}{(1-d^R_{st})}
		$$
	\end{enumerate}

	\item Update the Stick-breaking Weight $\pi_{k}^t$ and $q_k^t$:
	$$
	\pi_k^t \sim Beta(1+n_k^{t-1} + n_k^t, \gamma + \sum_{l = k+1}^K n_l^{t-1}+\sum_{l = k+1}^K n_l^t)
	$$
	$$
	n_k^{t-1} = \sum_{s=1}^{50}{\mathbb{I}(g^D[s, t-1]=k)} +  \sum_{s=1}^{50}{\mathbb{I}(g^R[s, t-1]=k)}
	$$
	$$
	n_k^t = \sum_{i=1}^N{(1-d^D_{st})}\mathbb{I}(g^D[st]=k) + \sum_{i=1}^N{(1-d^R_{st})}\mathbb{I}(g^R[st]=k) 
	$$
	$$
	q_k^t = \pi_k^t \prod_{l = 1}^{k-1}(1-\pi_l^t)
	$$
	\item Update $g^D[st]$ and $g^R[st]$\\
	Here we introduce the sampling algorithm for $g^D[st]$. The algorithm for $g^R[st]$ follows the same pattern. \\
	Let  $\mathbf{g}^D[t]\equiv (g^D[1t], g^D[2t], ..., g^D[50t])'$ and $\mathbf{q}^t\equiv (q_1^t, q_2^t, ..., q_K^t)'$. Let $Y^D_{st}$ represent the collection of $(V^D_{istj}, P^D_{istj}, S^D_{ist}, B^D_{ist})'$ for all $i= 1,2,..., I^D_{st}$ and $j = 1, 2, ..., J_t^V/J_t^P$;  $\mathbf{Y}_t^D \equiv (Y^D_{1t}, Y^D_{2t}, ..., Y^D_{50t})'$. Define $\Theta_k \equiv (\theta_k, \eta_k, \omega_k, \lambda_k)'$.\\
	We sample $\mathbf{g}^D[92]$,...,$\mathbf{g}^D[t]$, ..., $\mathbf{g}^D[73]$in turn. 
	$$
	Pr(g^D[st] = k | \mathbf{g}^D[92],...,\mathbf{g}^D[t+1], p,\mathbf{q}^{92},..., \mathbf{q}^{73} ,\mathbf{Y}^D_{92}, ... , \mathbf{Y}^D_{73}, \Theta_1, ..., \Theta_K) 
	$$
	$$
	\propto
	\underset{\text{part } 1}{\underbrace{ Pr(g^D[s,t+1]|p,g^D[st] = k, \mathbf{q}^{t+1})}}
	\underset{\text{part } 2}{\underbrace{Pr(g^D[st]=k|p, \mathbf{q}^{t},..., \mathbf{q}^{73} ,\mathbf{Y}^D_t, ... , \mathbf{Y}^D_{73},\Theta_1, ..., \Theta_K)}} 
	$$
	part 1:
	$$
	Pr(g^D[s,t+1] = l|p,g^D[st] = k, q_l^{t+1}) = (1-p)q^{t+1}_l + p\mathbb{I}(l=k)
	$$
	part 2:
	$$
	Pr(g^D[st]=k|p, \mathbf{q}^{t},..., \mathbf{q}^{73} ,\mathbf{Y}^D_t, ... , \mathbf{Y}^D_{73},\Theta_1, ..., \Theta_K))
	$$
	$$ \propto \underset{\text{part } a}{\underbrace{ f(Y_{st}| g^D[st]=k, \mathbf{Y}^D_{t-1}, ... , \mathbf{Y}^D_1, \Theta_1, ..., \Theta_K)}}
	\underset{\text{part } b}{\underbrace{Pr(g^D[st]=k|p, \mathbf{q}^{t},..., \mathbf{q}^{73} ,\mathbf{Y}^D_{t-1}, ... , \mathbf{Y}^D_{73},\Theta_1, ..., \Theta_K)}} 
	$$

	part a:
	$$
	f(Y^D_{st}| g^D[st]=k, \mathbf{Y}^D_{t-1}, ... , \mathbf{Y}^D_{73},\Theta_1, ..., \Theta_K) =  f(Y_{st}^D|\Theta_{g^D[st]})
	$$
	$$
	f(Y_{st}^D|\Theta_{g^D[st]}) = f_{V} f_{P} f_{S} f_{B}
	$$
	$$
	f_{V} = \prod_{i=1}^{I_{st}^D}\prod_{j=1}^{J_t^V}(\theta_{g^D[st]}^{V^D_{istj}}(1-\theta_{g^D[st]})^{(1 - {V^D_{istj}})})
	$$
	$$
	f_{P} = \prod_{i=1}^{I_{st}^D}\prod_{j=1}^{J_t^P}(\eta_{g^D[st]}^{p^D_{istj}}(1-\eta_{g^D[st]})^{(1 - {P^D_{istj}})})
	$$
	$$
	f_{S} = \prod_{i=1}^{I_{st}^D}(\omega_{g^D[st]}^{S^D_{ist}}(1-\omega_{g^D[st]})^{(1 - {S^D_{ist}})})
	$$
	$$
	f_{S} = \prod_{i=1}^{I_{st}^D}\frac{\lambda_{g^D[st]}^{B^D_{ist}}e^{-\lambda_{g^D[st]}}}{B^D_{ist}!}
	$$
	part b:
	$$
	Pr(g^D[st]=k|p,\mathbf{q}^{t},..., \mathbf{q}^{73} ,\mathbf{Y}^D_{t-1}, ... , \mathbf{Y}^D_{73},\Theta_1, ..., \Theta_K)
	$$
	$$
	= \sum_{l}Pr(g^D[st]=k|p,q^t_k, g^D[s,t-1]=l)Pr(g^D[s,t-1]=l|p,\mathbf{q}^{t-1},..., \mathbf{q}^{73} ,\mathbf{Y}^D_{t-1}, ... , \mathbf{Y}^D_{73},\Theta_1, ..., \Theta_K)
	$$
\end{enumerate}

\clearpage
\section{Simulation Study}\label{sec:sim}

In this section, we describe two simulations to highlight two kinds of group membership changes over time. In one simulation, we generate group memberships based on a  structural break model; in the other simulation, we generate group memberships based on a gradual  change model. We will show that the proposed method works well in both situations. 

\subsection{Data Generating Process}
 Recall the question of changing voting group we described in the motivating example. To simplify, we set up the question as the following: there are $T = 30$ parliamentary sessions, $N = 50$ representatives, and $M = 4$ issues to vote in each session. There are different voting groups in the parliament. For representative $i$ in a voting group $g$, the probability of voting ``yea'' for issue $j$ in session $t$ follows a Bernoulli distribution, $\mathsf{Bernoulli} (\theta_{gjt})$. We use different ways to generate group memberships in simulation 1 and simulation 2. 

\textbf{Simulation 1: a structural break model.} Two transition points at $t = 11 \text{\ and\ } 21$ separate the $30$ parliamentary sessions into three periods. In the first period, there are $3$ groups, with 20 representatives in group 1, 20 representatives in group 2, and 10 representatives in group 3; entering into the second period, 5 representatives in group 1 change to group 2 and 10 representatives in group 2 shift to group 1; in the last period, 5 representatives in group 1 change to group 2, 5 representatives in group 3 move to group 1, and the other 5 representatives in group 3 move to group 2. 

To summarize, there are 3 groups in the first and second periods, and only 2 groups in the last period. For each group, we generate the parameter $\theta_{gjt}$ of the Bernoulli distribution that models the voting outcomes from a uniform distribution. For group 1, the four uniform distributions for the four voting issues are $\mathbf{Uniform}(0.8, 1)$, $\mathbf{Uniform}(0.7, 1)$, $\mathbf{Uniform}(0, 0.2)$ and $\mathbf{Uniform}(0, 0.3)$; for group 2, they are $\mathbf{Uniform}(0, 0.2)$, $\mathbf{Uniform}(0, 0.3)$, $\mathbf{Uniform}(0.8, 1)$ and $\mathbf{Uniform}(0.7, 1)$; for group 3, they are $\mathbf{Uniform}(0.7, 1)$, $\mathbf{Uniform}(0, 0.2)$, $\mathbf{Uniform}(0, 0.3)$ and $\mathbf{Uniform}(0.8, 1)$. 

\textbf{Simulation 2: a gradual change model.} In the first 5 parliamentary sessions, there are  3 groups with 20, 20, and 10 representatives in each group. From $t = 6$ to $t = 25$, representatives in group 1 may change to group 2 with a probability of 0.5, and this change may happen at any time during this period; similarly, with a probability of 0.5, representatives in group 2 may shift to group 1 at a random time; for representatives in group 3,  they will shift to group 1 with a probability of 0.5 and otherwise they will shift to group 2. The process to generate $\theta_{gjt}$ for each group $g$ and each issue $j$ in session $t$ is the same as the process in simulation 1. 

\subsection{The Model}
In section \ref{model}, we only describe a general version of the proposed method. Here, we introduce the detailed model we use to analyze the simulated data. 

Let $g[it]$ represent the group of representative $i$ in session $t$. Then,
$$
g \sim \mathsf{IgCRP(\gamma, \alpha_p, \beta_p)}
$$
For $V_{ijt}$, the vote of issue $j$ that representative $i$ in sessions $t$ casts,  we assume it follows a Bernoulli distribution $\mathsf{Bernoulli(\theta_{jk})}$ for $g[it] = k$. Unlike the data generating process, we assume that $\theta_{jk}$ does not change with $t$. As we will show, the model still works well under this assumption. 
$$
V_{ijt} \sim \mathsf{Bernoulli(\theta_{j,g[it]})}
$$

For $g[it] = 1, 2, ..., k, ...$, we assume:
$$ 
\theta_{jk} \sim \mathsf{Beta}(\alpha_{
  \theta, \beta_{\theta}
}).
$$

The MCMC algorithm for this model is a simplified version of the MCMC algorithm we use for the empirical application. Thus, we skip the detailed algorithm here. 

\subsection{Results}
\begin{figure}[t]
  \centering
  \begin{tabular}{c}
   \begin{minipage}[t]{.5\textwidth}
    \centering
    \includegraphics[scale=.4]{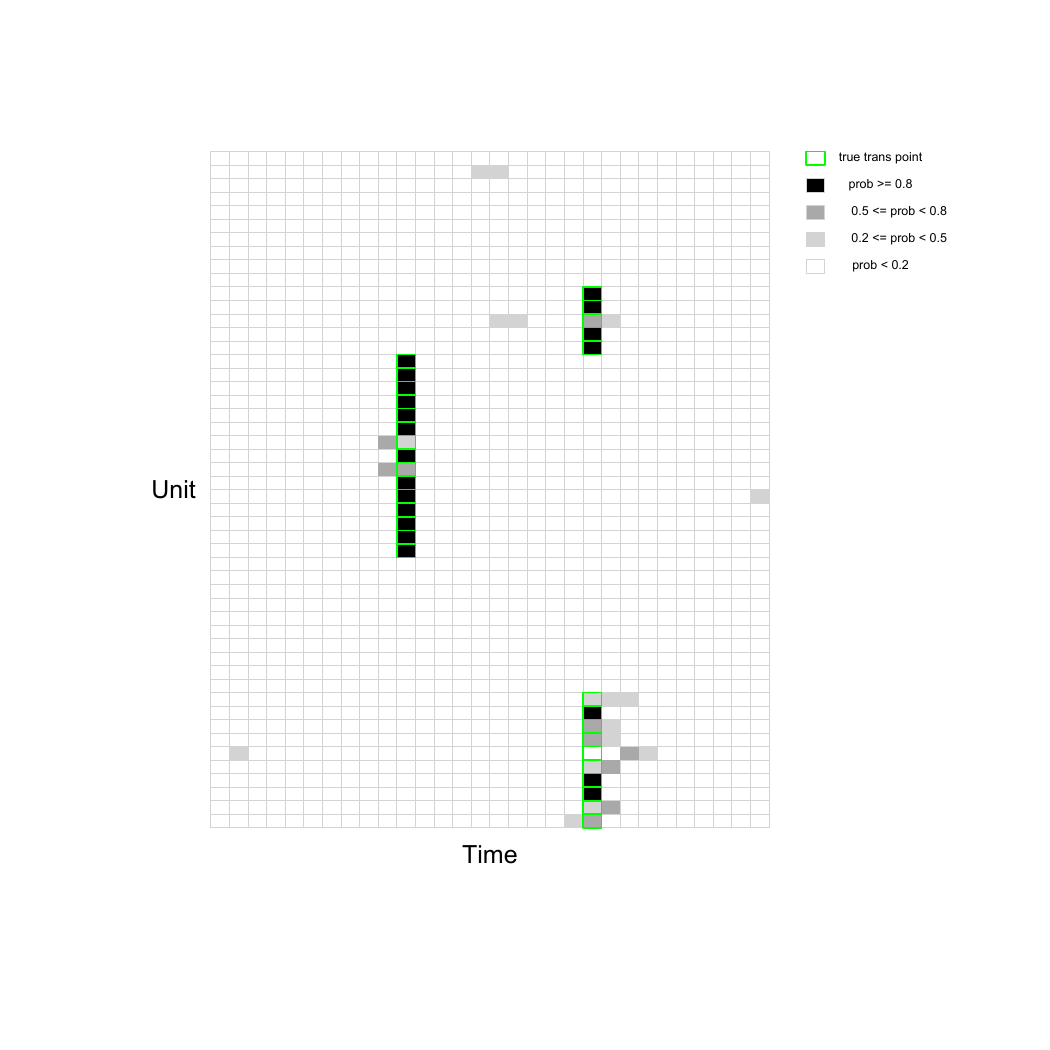}
    \newline
    \small (1) Recover Structural Break Model
   \end{minipage}
   
   \begin{minipage}[t]{.5\textwidth}
    \centering
    \includegraphics[scale=.4]{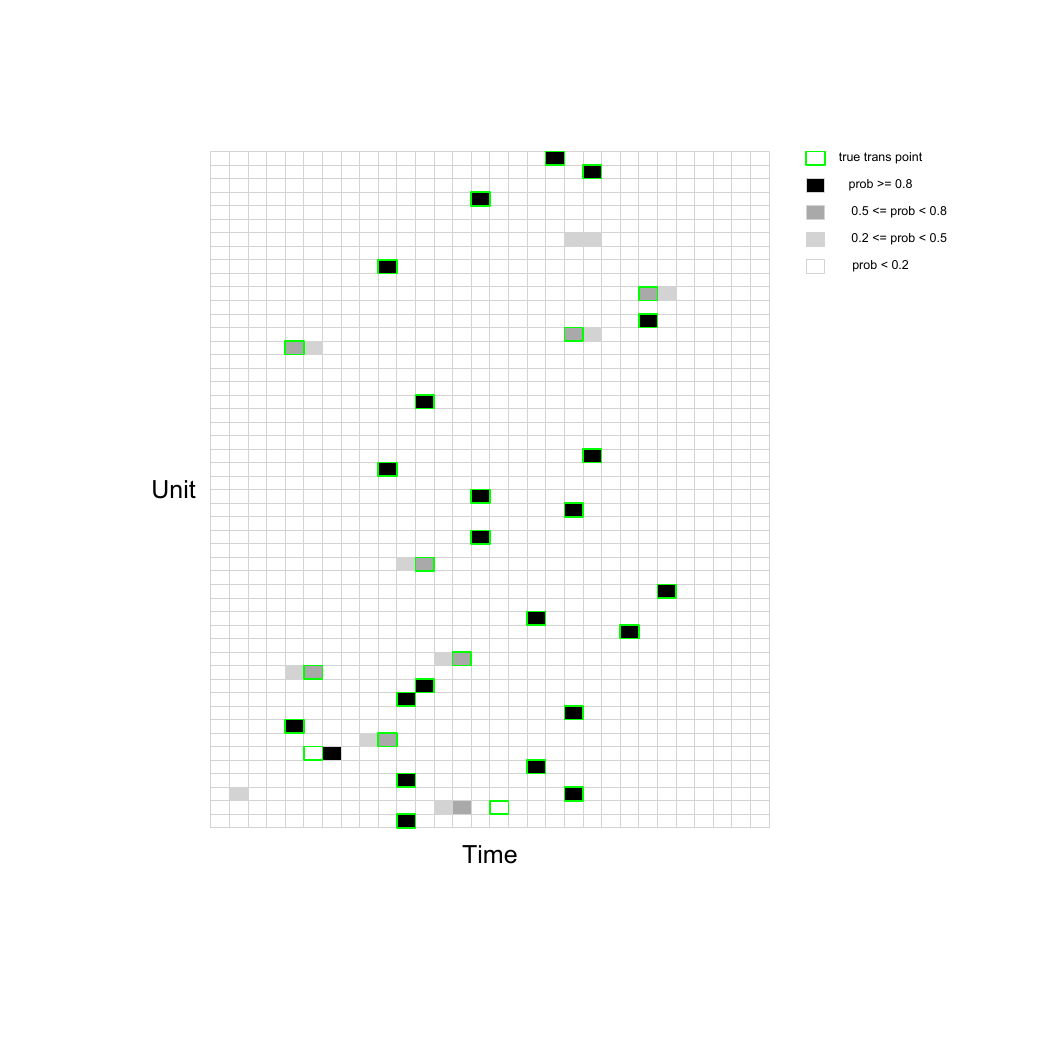}
    \newline
    \small (2) Recover Gradual Change Model
   \end{minipage}
  \end{tabular}
  \caption{Probabilities of Changing to a New Group. In the left figure, data is generated through a structural break model; in the right figure, data is generated through a gradual change model. A square represents the probability that the unit in the current time changes to a different group. The true transition points are circulated with green lines. This figure shows that no matter the data is generated through a structural break model or a gradual change model, the proposed method works well in identifying the transition point. }
  \label{fig:sim_point}
 \end{figure}

We first investigate whether the proposed method can detect the true transition points. For each unit, we calculate  the probability that the unit in the current time and in the former time are in different groups. A probability approaching 1 indicates a transition point.  As shown in Figure \ref{fig:sim_point}, the proposed method detects almost all transition points, successfully recovering both the structural break model and the gradual change model.  

\begin{figure}[t]
  \centering
  \begin{tabular}{c}
   \begin{minipage}[t]{.5\textwidth}
    \centering
    \includegraphics[scale=.4]{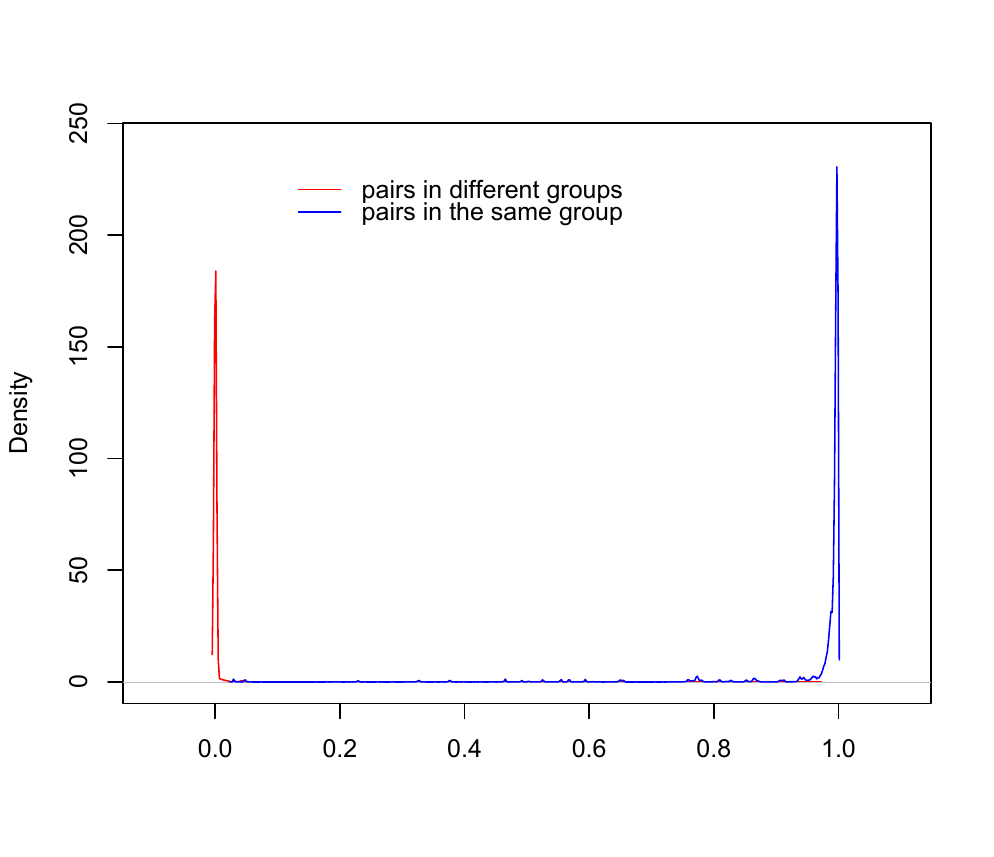}
    
    \small (1) Recover Structural Break Model
   \end{minipage}
   
   \begin{minipage}[t]{.5\textwidth}
    \centering
    \includegraphics[scale=.4]{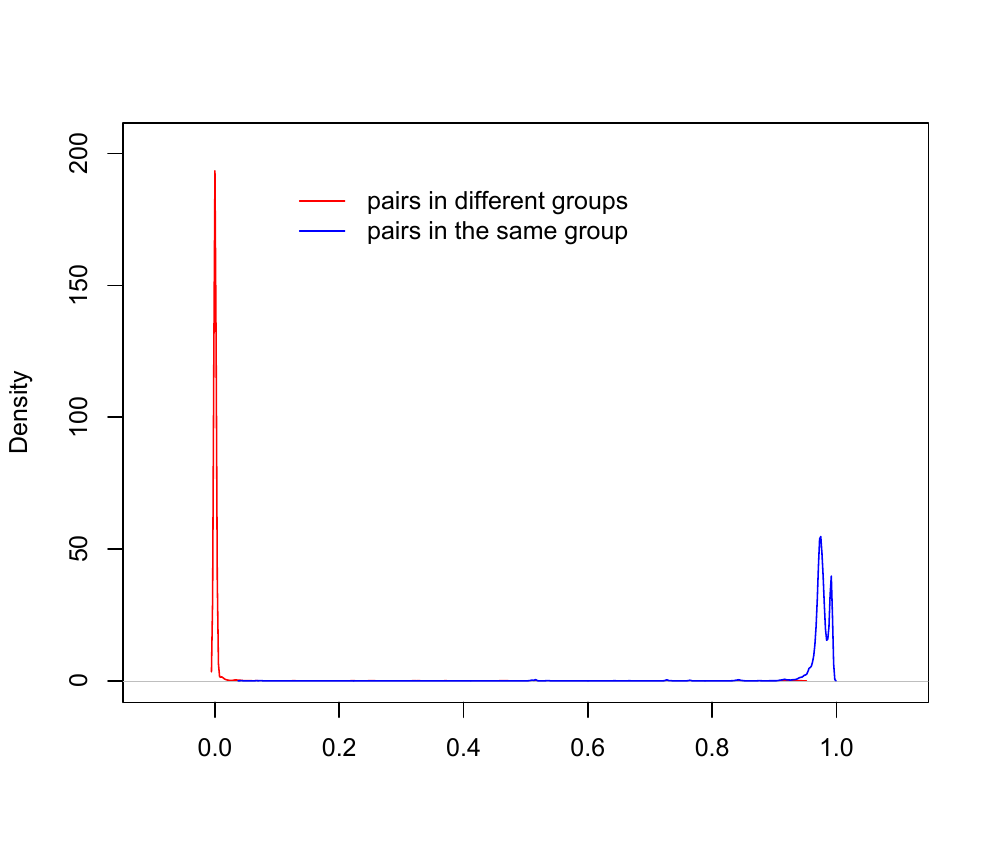}
    \newline
    \small  \ \ (2) Recover Gradual Change Model
   \end{minipage}
  \end{tabular}
  \caption{Densities of Probabilities that Two Observations are in the Same Groups. As we know the true memberships, we can separate pairs in the same groups from pairs in different groups and calculate the densities separately.}
  \label{fig:density}
 \end{figure}

Besides checking whether the proposed method can detect the true transition points, we also investigate whether our method recovers true group memberships across units and over time together. For this purpose, we calculate the probability that two observations (they are either from different units,or in different time points, or both) are in the same group for all possible pairs. As we know the true group memberships in simulation studies, we separate pairs in different groups from pairs in the same group. For pairs in different groups, we expect the density of probabilities that two observations are in the same group to center around 0; for pairs in the same, we expect the density of probabilities to center around 1. As shown in Figure \ref{fig:density}, the proposed method discovers the true group memberships for both the structural break model and the gradual transition model.

\end{document}